\documentclass[a4paper]{article}

\usepackage{latexsym}
\usepackage{graphics}
\usepackage{epsfig,hangcaption}
\usepackage{multirow}
\usepackage{amssymb,amsmath}
\usepackage{times}

\newcommand{\shat}{\mbox{\ensuremath{\hat{s}}}}

\newcommand{\Mfunction}[1]{\mathbf{#1}}

\DeclareOldFontCommand{\tt}{\normalfont\sf\small}{\mathtt}

\newcommand{\authorrunning}[1]{\edef\authorrun{#1}}
\newcommand{\titlerunning}[1]{\edef\titlerun{#1}}
\usepackage{fancyhdr}
\begin{document}

%

\title{Improved Phase Space Treatment of Massive Multi-Particle Final States}

\titlerunning{Improved Phase Space Treatment\ldots}

\authorrunning{{B. P. Kersevan,E. Richter-Was}}

\author{
Borut Paul Kersevan \\
\small Faculty of Mathematics and Physics, University of Ljubljana,
 Jadranska 19, SI-1000 Ljubljana, Slovenia. \\ \small and  \\
\small  Jozef Stefan Institute, Jamova 39, SI-1000 Ljubljana, Slovenia.\\[10pt]
Elzbieta Richter-W\c{a}s\thanks{ \scriptsize 
Partly supported by Marie Curie Host Fellowship for the Transfer of 
  Knowledge Contract No. MTKD-CT-2004-510126  and  
  by the  EC FP5 Centre of Excellence ``COPIRA'' under the contract 
  No. IST-2001-37259.}\\
\small Institute of Physics, Jagellonian University,\\
\small 30-059 Krakow, ul. Reymonta 4, Poland;\\ \small and  \\
\small Institute of Nuclear Physics PAS,  31-342 Krakow, ul. Radzikowskiego 152, 
Poland.
}

\maketitle

\begin{picture}(0,0)
\put(350,330){\rm   TPJU-5/2004}\\
\put(350,340){\rm   hep-ph/0405248}\\
\end{picture}

\abstract{\it 
In this paper the revised Kajantie-Byckling approach and improved phase space
sampling techniques for the massive multi-particle final states are
presented. The application of the developed procedures to the processes
representative for LHC physics indicates the possibility of a substantial
simplification of multi-particle phase space sampling while retaining a
respectable weight variance reduction and unweighing efficiencies in the event
generation process.
\bigskip \hrule}

\begin{center}
{\bf Key Words:} Phase space generation, SM backgrounds at LHC, 
massive final state particles, Kajantie-Byckling approach, 
Monte Carlo generator, heavy flavor production,  
multi-channel phase-space generation\\
{\bf PACS:} 02.60.Cb 02.70.Lq 13.38.-b , 13.90.+i 
\end{center}


\section{Introduction}

With the advent of LHC era a need for precise predictions (and subsequently
detailed simulation) of many QCD and electroweak processes has arisen. One of
the necessary components of an accomplished simulation tool is definitely
efficient phase space modeling of multi-particle final states. While the light
(massless) particles in the final states and the relevant topologies can quite
effectively be simulated using general tools (e.g. {\tt RAMBO,SARGE,HAAG} 
\cite{rambo,sarge,haag}) further development might be needed for phase space
sampling where the massive final state particles are present. While at LEP's 200
GeV a massless approximation for final state particles was adequate, in most of
the studied cases at LHC in contrast the massless approximation becomes less
suitable because of crossing the top quark production barrier in conjunction
with the possibility of multi-quark final states and the shifting centre-of-mass
energy in proton-proton collisions. Furthermore, in the LEP era the number of
(hard process) particles in the final states of relevance only rarely rose above
four and subsequently the number of possible final state topologies was
relatively modest. Typical processes of interest (signal) and their backgrounds
at LHC will involve several heavy quarks and/or weak bosons in the intermediate
states; thus the number of particles in a typical process under study is
generally at least four. Furthermore, the number of Feynman diagrams
contributing to a typical process involving heavy quarks in the final state is
mostly ranging from a few to hundred(s) (and can steeply rise to many thousands
when massless quarks are added). Subsequently, this results in varied particle
topologies which prove to be a challenge when trying to adequately describe them
using the available statistical approaches and numerical methods.  Important
steps have already been made in matrix element calculations (e.g. MadGraph\cite{madgraph}) 
and have so far surpassed the corresponding development of phase-space modeling and
sampling techniques.

The general objective in simulation of physics processes for the LHC environment
is thus to improve the integration of the differential cross-section using
Monte-Carlo sampling methods\footnote{For a nice discussion on the topic see
e.g. \cite{jadach,was}\ldots}. The sampling method used should aim to minimise
the variance of the integral as well as maximise the sampling efficiency given a
certain number of iterations and the construction of the sampling method itself
should aim to be sufficiently general and/or modular to be applicable to a wide
range of processes. Writing down a (process) cross-section integral for LHC type
(hadron-hadron) collisions:
\begin{equation}
\sigma = \int \sum_{a,b} f_a(x_1,Q^2) f_b(x_2,Q^2) \frac{|{\mathcal{M}_n}|^2}{(2 \pi)^{3n-4} 
(2\hat{s})}\, dx_1\, dx_2\, d\Phi_n,
\end{equation}
where $\rm f_{a,b}(x,Q^2)$ represent the gluon or (anti)quark parton density
functions, $\rm |{\mathcal{M}_n}|^2$ the squared n-particle matrix element divided
by the flux factor $\rm [(2 \pi)^{3n-4} 2\hat{s}]$ and $d\Phi_n$ denotes the
n-particle phase space differential. The quantity $\rm \hat{s} = x_1\, x_2\, s$ is
the effective centre-of-mass energy, and the sum $\rm \sum_{a,b}$ runs in case of
quark-antiquark incident partons over all possible quark-antiquark combinations
($\rm a,b = u,d,s,c,\bar{u},\bar{d},\bar{s},\bar{c}$). In case of $\rm g g$ initial
state the sum has only one term with $\rm a=b=g$.

It is often  convenient to re-write the differential cross-section in the
form:
\begin{equation}
\sigma = \int \sum_{a,b} x_1 f_a(x_1,Q^2) \; x_2 f_b(x_2,Q^2) \frac{|{\mathcal{M}_{n}}|^2}
{(2 \pi)^{3n-4} (2 \hat{s}^2)}\, dy\, d\hat{s}\, d\Phi_{n},
\label{e:dsig}
\end{equation}

with the new (rapidity) variable given by $\rm y = 0.5 \log(x_1/x_2)$. 
The n-body phase-space differential $\rm d\Phi_n$ and its integral $\Phi_n$
depend only on $\hat{s}$ and particle masses $m_i$  due to Lorentz invariance:
\begin{equation}
\Phi_n(\hat{s},m_1,m_2,\ldots,m_n) = \int d \Phi_n(\hat{s},m_1,m_2,\ldots,m_n) = 
\int \delta^4\left((p_a + p_b) - \sum_{i=1}^n p_i\right) \prod_{i=1}^n d^4p_i \delta (p_i^2 - m_i^2) \Theta(p_i^0),
\label{e:phins}
\end{equation}
with $\rm a$ and $\rm b$ denoting the incident particles and $\rm i$ running
over all outgoing particles $i=1,\ldots,n$. What one would like to do is to
split the n-body phase parameterised by 3n-4 essential (i.e. non-trivial)
independent variables into manageable subsets (modules) to be handled by
techniques which reduce the variance of the result and/or the sampling
efficiency (e.g. importance sampling\cite{kleiss} or adaptive integration 
like VEGAS\cite{vegas} or FOAM\cite{foam}). Stating this in formal terms, 
the above Equation \ref{e:dsig} should be transformed into an expression like:
\begin{equation}
\sigma = \left(\prod_{i=1}^n \int\limits_{s_{i}^{-}}^{s_{i}^{+}} ds_i \right) 
\left(\prod_{j=1}^m \int\limits_{t_{j}^{-}}^{t_{j}^{+}} dt_j \right) 
\left(\prod_{k=1}^l \int\limits_{\Omega_{k}^{-}}^{\Omega_{k}^{+}} d\Omega_k \right) 
\left|\mathcal{J}_n\right| \int \sum_{a,b} x_1 f_a(x_1,Q^2) \; x_2 f_b(x_2,Q^2) \frac{|{\mathcal{M}_{n}}|^2}
{(2 \pi)^{3n-4} (2 \hat{s}^2)}\, dy\, d\hat{s}\,  
\label{e:dsigmod}
\end{equation}
where one integrates over Mandelstam type (Lorentz invariant) momenta transfers
$\rm s_i,t_j$ and space angles $\Omega_k \equiv (\cos \vartheta_k, \phi_k) $
 within the kinematically allowed limits (3n-4 variables in total) with the term $\rm
|\mathcal{J}_n|$ denoting the Jacobian of the transformation.  If one would then
decide to introduce importance sampling functions in order to reduce the peaking
behavior of the integrand \cite{kleiss}, the integrals would take the form:
\begin{equation}
 \int\limits_{s_{i}^{-}}^{s_{i}^{+}} ds_i = 
\int\limits_{s_{i}^{-}}^{s_{i}^{+}} \frac{g_i(s_i)}{g_i(s_i)} ds_i,
\end{equation}
where the importance sampling function $\rm g_i$ is  probability density function
normalised in the integration region $\rm [s_{i}^{-},s_{i}^{+}]$:
\begin{equation}
\int\limits_{s_{i}^{-}}^{s_{i}^{+}}{g_i(s_i)} ds_i =1,
\end{equation}
which exhibits a similar peaking behavior as the integrand. Formally, one then inserts the 
identity:
\begin{equation}
  1 = \int\limits_0^1 \delta\left(r_i - \int\limits_{s_i^{-}}^{s_{i}} g_i(s_i) ds_i \right)  dr_i
\end{equation}
into the integral and then derives the \emph{unitary} sampling prescription: 
\begin{equation}
  \int\limits_0^1 dr_i \int\limits_{s_{i}^{-}}^{s_{i}^{+}} 
\delta\left(r_i - \int\limits_{s_i^{-}}^{s_i} g_i(s_i) ds_i \right)  \frac{g_i(s_i)}{g_i(s_i)} ds_i = 
\int\limits_0^1 dr_i \int\limits_{s_{i}^{-}}^{s_{i}^{+}} 
\delta\left(s_i - G^{-1}(r_i)\right)  \frac{1}{g_i(s_i)} ds_i = \int\limits_0^1
\frac{dr_i}{g_i(G^{-1}(r_i))},
\label{e:usampling}
\end{equation}
which formally means that the $\rm s_i$ values are sampled from the interval
according to the $\rm g_i(s_i)$ distribution by using \mbox{(pseudo-)random}
variable $\rm r_i$ together with the $\rm g_i(s_i)$ cumulant
$G(s_i)=\int_{s_i^{-}}^{s_{i}} g_i(s_i) ds_i$ with its inverse $G^{-1}$. The
unitarity of the algorithm states that each trial ( $\rm r_i$ value) produces a
result (i.e. a corresponding $\rm s_i$ value distributed according to $\rm g_i(s_i)$). 

Performing such substitutions on all integration parameters would give as the cross-section
expression;
\begin{equation}
\sigma = \prod_{i=1}^{3n-4} \int\limits_0^1 dr_i \frac{f(r_1,r_2\ldots)}{g(r_1,r_2\ldots)}
\end{equation}
where the integrand would (hopefully) have as low variation as possible at least for a subset
of contributing Feynman diagrams\footnote{The 'modularisation' can be performed for several 
topologies at the same time and multi-channel techniques can be applied.}. To improve the sampling 
method further, the $\rm r_i$ (pseudo-)random variables can be sampled from adaptive algorithms of the VEGAS 
type \cite{vegas,acermc}. 

\begin{figure}[ht]
\begin{center}
     \epsfig{file=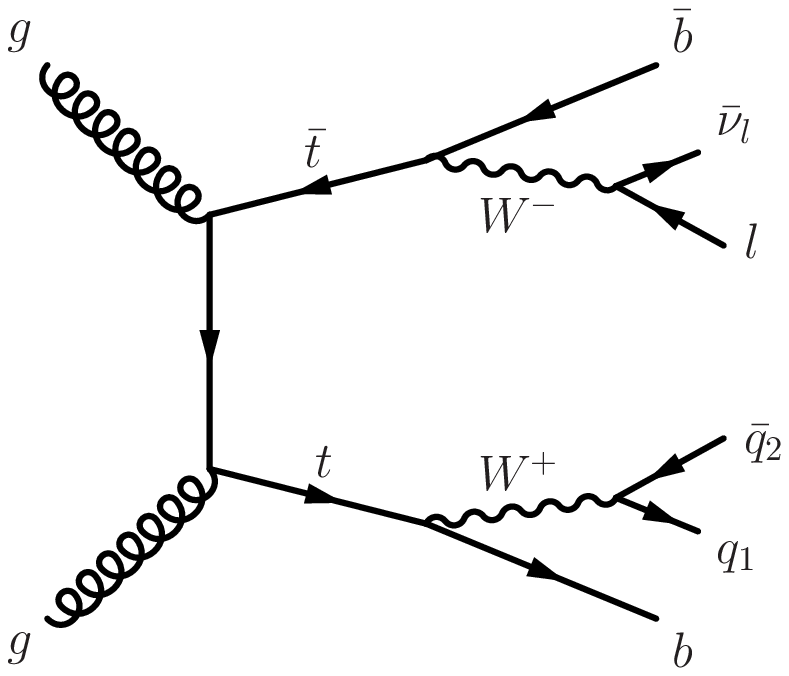,width=4.0cm}
     \epsfig{file=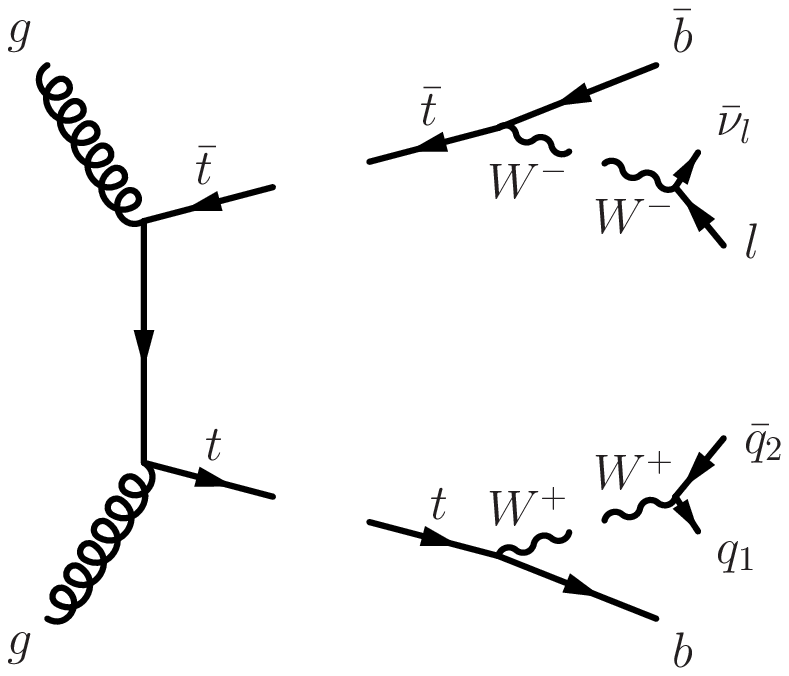,width=4.0cm}
     \epsfig{file=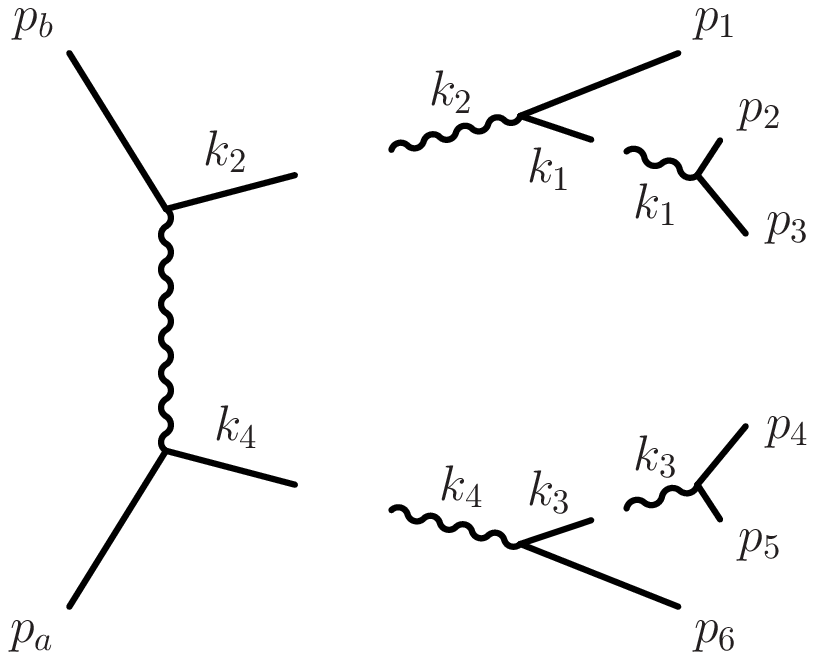,width=4.0cm}
\end{center}
\isucaption{
\small A representative Feynman diagram describing a $2 \to 6$ process $gg \to t\bar{t} \to b \bar{b} W^+ W^- \to b \bar{b} \ell \bar{\nu}_\ell q_1 \bar{q}_2$ and its decomposition into a set of $2 \to 2$ t-channel and s-channel sub-processes.
\label{f:ttbar}} 
\end{figure}

A representative Feynman diagram describing a $2 \to 6$ process is shown in
Figure \ref{f:ttbar}. As one can see, the process can be split in
several consecutive branchings, this approximation is often used in
matrix element (probability amplitude) calculations. It seems rather
obvious that any Feynman diagram can be split in a series of
horizontal and vertical branchings that one can denote as s-type and
t-type(u-type) using the analogy with the Mandelstam variables. What
one would like to do is thus to modularise the phase space in the form
of sequential s- and t-type splits.

The s-splitting of phase space is relatively easy to do and has as such been
used in many instances of Monte--Carlo generation (e.g. {\tt FermiSV}  \cite{fermisv}, 
{\tt Excalibur} \cite{excal}, {\tt Tauola} \cite{tauola} etc..); the t-type
branchings (often tagged as multi-(peri)pheral topologies) have in contrast
generally been calculated only for specific cases (e.g. for 3 or 4 particles in
the final state \cite{fermisv,excal}). As it turns out, the problem of several
massive particles in the final state has already appeared more than 30 years ago
when several hadrons (e.g. pions) have been produced in (comparatively low
energy) nuclear interactions. At that time Kajantie and Byckling \cite{KB} have
derived the formulae for simulating any sequence of s- and t- type branchings
which, with some modifications, can also be applied to the EW and QCD processes
involving heavy quarks and/or massive bosons at LHC.

In the following Sections \ref{s:kb} and \ref{s:prop} the revised
version of of Kajantie-Byckling algorithm (KB) will be discussed and a
formulation of the algorithm for the multi-phase space integration
will be presented. In Section \ref{s:num2} the numerical results for
some representative applications as implemented in the {\tt AcerMC
2.0} Monte--Carlo generator will be presented.
\clearpage 

\section{Modified Kajantie-Byckling Formalism\label{s:kb}}

\subsection{The s-type Branching Algorithms}
\begin{figure}[htb]
\begin{center}
     \epsfig{file=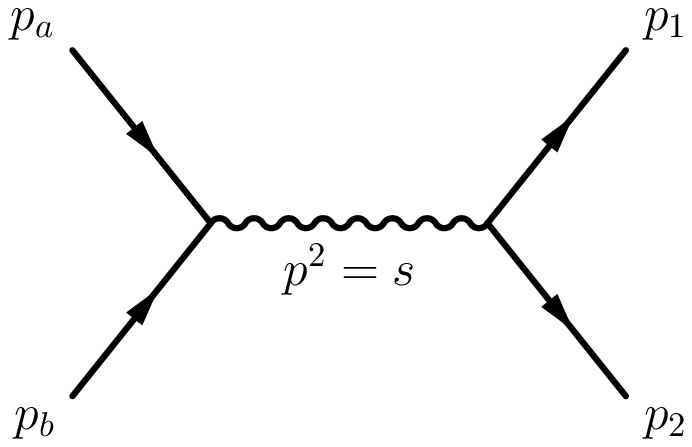,width=3.5cm}
\end{center}
\isucaption{
\small A diagram of a generic $\rm 2 \to 2$ s-channel process.
\label{f:schan}}
\end{figure}

The s-splits are the simplest method in the KB formalism. For the sake
of completeness one should start with the definition of the two-body
phase space integral (c.f. Fig \ref{f:schan}):
\begin{equation}
\Phi_2(s,m_1,m_2)=\int d^4p_1 d^4p_2 \delta (p_1^2 - m_1^2) \delta
(p_2^2 - m_2^2) \delta^4 (p - p_1 -p_2) \Theta(p_1^0) \Theta(p_2^0)  ,
\end{equation}
with the incoming momentum sum $\rm p = (p_a + p_b), p^2 = s$ and outgoing
momenta $\rm p_{1,2}, p_{1,2}^2 = m_{1,2}^2 $. The phase space integral is
Lorentz invariant (as one can observe in the above Equation where it is written
in a manifestly Lorentz invariant form). Subsequently, due to Lorentz
invariance, the integral is necessarily a function of the Lorentz scalars $\rm
s, m_1$ and $\rm m_2$ only. The step function product $\rm \Theta(p_1^0)
\Theta(p_2^0)$ is the explicit requirement of the positiveness of the energy
terms in $\rm p_{1,2}$ while the delta functions represent the on-shell
conditions on $\rm p_{1,2}$ and the total momentum conservation.

The integral can be transformed into a more compact form by
integrating out the spurious variables; one thus first integrates over
$\rm d^4 p_2$ and chooses the centre-of-mass system (CMS) as the
integration system of reference with $\rm p = (\sqrt{s},0,0,0)$ and
then evaluates the integrals over $\rm p_1^0$ and $\rm E_1^*$:
\begin{eqnarray}
\Phi_2(s,m_1,m_2) &=& \int  d^4p_1 \delta (p_1^2 - m_1^2) \delta
((p-p_1)^2 - m_2^2)  \Theta(p_1^0) \\ \notag
&=& \int \frac{d^3p_1^*}{2 E_1^*} \delta (s + m_1^2 - 2\sqrt{s} E_1^* - m_2^2)\\ \notag
&=& \frac{1}{4 \sqrt{s}} \int p^1_* dE_1^* d\Omega_1^* \delta \left( E_1^* - \frac{s +
m_1^2 -m_2^2}{2 \sqrt{s}} \right)\\ \notag
&=&\frac{p_1^*(s,m_1,m_2)}{4 \sqrt{s}} \int d\Omega_1^*,
\label{e:phi2s}
\end{eqnarray}
with the stars explicitly denoting the values in the centre-of mass system. 
The first integration simply sets $\rm p_1^0 = \sqrt{ (p_1^*)^2 + m_1^2 } =
E_1^*$ and the second integral leads to the well known relations for the energy:
\begin{equation}
E_1^*= \frac{s + m_1^2 -m_2^2}{2 \sqrt{s}}, ~~~~ E_2^* = \sqrt{s}- E_1^* = 
\frac{s + m_2^2 -m_1^2}{2 \sqrt{s}},
\label{e:cmse}
\end{equation}
and momenta sizes: 
\begin{equation}
p_1^*= |\vec{p}_1^*|=\frac{\sqrt{\lambda(s,m_1^2,m_2^2)}}{2\sqrt{s}}, ~~~~ 
p_2^*= p_1^*
\label{e:cmsp}
\end{equation}
of two particle production. The $\rm \lambda(s,m_1^2,m_2^2)$ denotes the Lorentz
invariant function:
\begin{equation}
\lambda(s,m_1^2,m_2^2)=(s-(m_1+m_2)^2)(s-(m_1-m_2)^2)
\label{e:lambda}
\end{equation}
and thus explicitly contains the phase space cutoff, i.e. the requirement that
the available CMS energy $\rm \sqrt{s}$ should be bigger than the mass sum $\rm
\sqrt{s} \geq (m_1+m_2)$. Note that the integration was so far done only over
the spurious parameters, leaving the polar and azimuthal angle of the $\rm p_1$
particle as the two independent parameters $\rm d\Omega^* = d\cos\theta^*
d\varphi^*$.  The integral becomes trivial to sample in case the outgoing
particles can be approximated as massless (the 'boost' factor lambda transforms
to unity). As already claimed, the latter approximation is however often
unjustified when studying processes representative for the LHC environment.

\begin{figure}[ht]
\begin{center}
     \epsfig{file=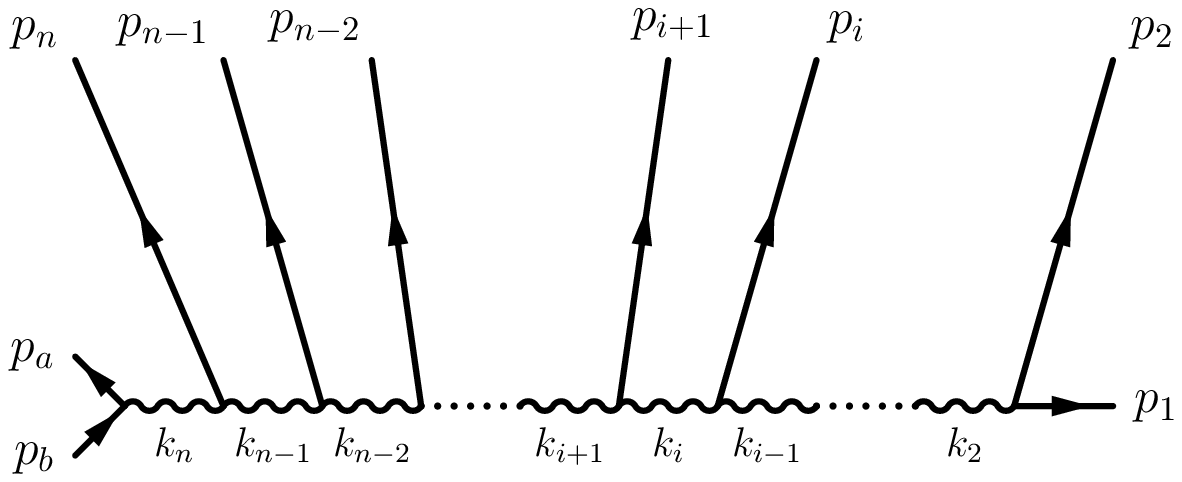,width=8.0cm}
\end{center}
\isucaption{
\small The diagrammatic representation of consecutive s-splits.
\label{f:ssplits}} 
\end{figure}

Kajantie and Byckling \cite{KB} introduced the \emph{recursion} and \emph{splitting} relations
for the n-particle phase space $\rm \Phi_n(s)$ given by Eq. \ref{e:phins}. The
recursion relation can be derived by defining the momentum sum:
\begin{equation}
k_i = \sum_{j=1}^{i} p_j = (k_i^0,\vec{k_i}) ;~~~ M_i^2 = k_i^2.
\label{e:kidef}
\end{equation}
Subsequently one can interpret $\rm p = k_n$ and $ s = M_n^2$ from Eq. \ref{e:phins}.
One continues by introducing the identities:
\begin{equation}
1 = \int dM_{n-1}^2 \delta(k_{n-1}^2 - M_{n-1}^2) \Theta(k_{n-1}^0)
\label{e:rec1}
\end{equation}
and
\begin{equation}
1 = \int d^4k_{n-1} \delta^4(p - k_{n-1} - p_n)
\label{e:rec2}
\end{equation}
into the integral of Equation \ref{e:phins}; separating out the arguments
containing $\rm k_{n-1}$ and $p_n$ terms one obtains:
\begin{eqnarray}
\Phi_n(M_n^2,m_1,m_2,\ldots,m_n) &=& \int dM_{n-1}^2 \times\\ \notag
&\times&\left\{ \int d^4k_{n-1} d^4p_n \delta(k_{n-1}^2 - M_{n-1}^2) \delta (p_n^2 - m_n^2)
\delta^4(p - k_{n-1} - p_n)\Theta(k_{n-1}^0)\Theta(p_{n}^0) \right\}\times\\ \notag
&\times& \Phi_{n-1}(M_{n-1}^2,m_1,m_2,\ldots,m_{n-1}),
\end{eqnarray}
where the remaining $p_i$ terms form the (n-1)-particle phase space integral
$\rm \Phi_{n-1}(M_{n-1}^2,m_1,m_2,\ldots,m_{n-1})$ and the terms in curly
brackets give a two particle phase space term (c.f. Eq. \ref{e:phi2s}):
\begin{eqnarray}
\Phi_n(M_n^2,m_1,m_2,\ldots,m_n) &=& \int dM_{n-1}^2 \Phi_2(M_n^2,M_{n-1},m_n) 
 \Phi_{n-1}(M_{n-1}^2,m_1,m_2,\ldots,m_{n-1}) \label{e:recur}
\\ \notag
&=&  \int dM_{n-1}^2 \frac{p_n^*}{4 M_n}
\Phi_{n-1}(M_{n-1}^2,m_1,m_2,\ldots,m_{n-1}) \\ \notag
&=& \int\limits_{(\sum_{i=1}^{n-1}m_i)^2}^{(M_n - m_n)^2} dM_{n-1}^2 
\frac{\sqrt{\lambda(M_n^2,M_{n-1}^2,m_n^2)}}{8 M_n^2} \int d\Omega_n^*
 \Phi_{n-1}(M_{n-1}^2,m_1,m_2,\ldots,m_{n-1}),
\end{eqnarray}
with the integration limits on $\rm M_{n-1}^2$ following from its definition in
Eq. \ref{e:kidef}. It has to be emphasized that the angles in $d\Omega_i^*$ are
each time calculated in the centre-of-mass system of $\rm k_i$ with the
invariant mass $\rm M_i$. The resulting recursion relation is clearly of
advantage when describing cascade decays of particles $k_n \to k_{n-1} p_n \to
k_{n-2}, p_n, p_{n-1} \to \ldots$; it also proves that the n-particle phase space
of Eq. \ref{e:phins} can be reduced into a sequence of two-particle phase space
terms, as shown in Figure \ref{f:ssplits}.
 
It can further prove of advantage to loosen up the splitting terms of
Eqns. \ref{e:rec1},\ref{e:rec2} so that instead of summing to n-1 one groups an
arbitrary set of $\rm \ell$ particles:
\begin{eqnarray}
1 &=& \int dM_{l}^2 \delta(k_{l}^2 - M_{l}^2) \Theta(k_{l}^0), \\ 
1 &=& \int d^4k_{l} \delta^4(p - k_{l} - \sum_{j=l+1}^{n} p_j),
\label{e:split}
\end{eqnarray}
which, when repeating the procedure in recursion relation of Eq. \ref{e:recur},
results in an expression:
\begin{equation}
\Phi_n(M_n^2,m_1,m_2,\ldots,m_n) = 
\int\limits_{(\sum_{i=1}^{l} m_i)^2}^{(M_{l+1} - m_{l+1})^2} dM_{l}^2 
\Phi_{n-l+1}(M_n^2,M_l,m_{l+1},\ldots,m_n) \Phi_{l}(M_l^2,m_1,m_2,\ldots,m_l),
\end{equation}
and thus effectively splits the phase space into two subsets, equivalent to
introducing an intermediate(virtual) particle with momentum $k_l$. 

The number of splitting relations and the number of particles in each group as
given in Eq. \ref{e:split} can be chosen in any possible sequence, thus meaning
that the grouping sequence is arbitrary and can be adjusted to fit the topology
in question.\footnote{Suggestions of \cite{KB} on how to pick random number
sequences will not be used since one might like to couple this method with an
adaptive algorithm to improve the sampling efficiencies.}

At this point some modifications were introduced to the algorithm in order to
adapt it to the specifics of the processes expected at the  LHC.
Kajantie and Byckling namely assumed that the generation sequence would be 'down' the
cascade (i.e. by sampling first a $\rm M_n$ value, then $M_{n-1}$ value
etc\ldots~ as is indeed most often done in Monte-Carlo Generators). This might
however not be optimal in the LHC environment since the available centre-of-mass
energy for the hard process (\shat)\; can vary in a wide range of values
(c.f. Equation \ref{e:dsig}) and has to be sampled from a distribution
itself. The shape of the distribution function for \shat\; is expected to behave
as a convolution of the peaking behavior of all participating invariant masses
times the parton density functions (c.f. Eq. \ref{e:dsig}); it subsequently
seems to be more natural (and efficient) first to sample the individual
propagator peaks and then their subsequent convolutions. Furthermore, by
generating the invariant masses 'up' the cascade (i.e. first $\rm M_2$, 
$\rm M_3$ \ldots $\rm M_n$ and finally $\rm \hat{s}$) the kinematic limits on
the branchings occur in a more efficient way (bound on the $\sqrt{\lambda}$
values, see Equations \ref{e:lambda} and \ref{e:revlim}), which is very
convenient since in the LHC environment no stringent generation cuts should be
made on the inherently non-measurable $\rm \hat{s}$  as it cannot be
accounted for by an analogous cut in a physics analysis.

A necessary modification of the algorithm would thus be to reverse the
generation steps by starting with the last pair(s) of particles. In terms of
integration (i.e. sampling) limits this translates into changing the limits of
Eq. \ref{e:recur}:
\begin{eqnarray}
\Phi_{n}(M_n^2,m_1,m_2,\ldots,m_n) &=& \\  \notag
&=& \int\limits_{(\sum_{i=1}^{n-1} m_i)^2}^{(M_n - m_n)^2} dM_{n-1}^2 
\frac{\sqrt{\lambda(M_n^2,M_{n-1}^2,m_n^2)}}{8 M_n^2} \int d\Omega_n^*\\ \notag &\times&
 \int\limits_{(\sum_{i=1}^{n-2} m_i)^2}^{(M_{n-1} - m_{n-1})^2} dM_{n-2}^2 
\frac{\sqrt{\lambda(M_{n-1}^2,M_{n-2}^2,m_{n-1}^2)}}{8 M_{n-1}^2} \int d\Omega_{n-1}^*
\\ \notag &\times& \ldots  
 \int\limits_{(\sum_{j=1}^{i-1} m_j)^2}^{(M_{i} - m_{i})^2} dM_{i-1}^2 
\frac{\sqrt{\lambda(M_{i}^2,M_{i-1}^2,m_{i}^2)}}{8 M_{i}^2} \int d\Omega_{i}^*
\ldots  \\ \notag &\times&
\int\limits_{(m_1 +m_2)^2}^{(M_3 - m_{3})^2} dM_{2}^2 
\frac{\sqrt{\lambda(M_{3}^2,M_{2}^2,m_3^2)}}{8 M_{3}^2} \int d\Omega_{3}^*
\\ \notag &\times&
\frac{\sqrt{\lambda(M_{2}^2,m_1^2,m_2^2)}}{8 M_{2}^2} \int d\Omega_{2}^*,
\end{eqnarray}
which accommodates the mass generation sequence: $k_n \to k_{n-1} + p_n \to
\ldots$ (i.e. first sample $M_{n-1}^2$, then $M_{n-2}$ etc\ldots), into 
\begin{eqnarray}
\Phi_{n}(M_n^2,m_1,m_2,\ldots,m_n) &=& \label{e:revlim} \int\limits_{(M_{n-2}+m_{n-1})^2}^{(M_n - m_n)^2} dM_{n-1}^2 
\frac{\sqrt{\lambda(M_n^2,M_{n-1}^2,m_n^2)}}{8 M_n^2} \int d\Omega_n^* \\ \notag &\times&
 \int\limits_{(M_{n-3}+m_{n-2})^2}^{(M_n - m_n - m_{n-1})^2} dM_{n-2}^2 
\frac{\sqrt{\lambda(M_{n-1}^2,M_{n-2}^2,m_{n-1}^2)}}{8 M_{n-1}^2} \int d\Omega_{n-1}^*
\\ \notag &\times& \ldots  
 \int\limits_{(M_{i-1} + m_i)^2}^{(M_n - \sum_{j=i+1}^{n} m_{j})^2} dM_{i-1}^2 
\frac{\sqrt{\lambda(M_{i}^2,M_{i-1}^2,m_{i}^2)}}{8 M_{i}^2} \int d\Omega_{i}^*
\ldots  \\ \notag &\times&
\int\limits_{(m_1 +m_2)^2}^{(M_n - \sum_{j=3}^{n} m_{j})^2} dM_{2}^2 
\frac{\sqrt{\lambda(M_{3}^2,M_{2}^2,m_3^2)}}{8 M_{3}^2} \int d\Omega_{3}^*
\\ \notag &\times&
\frac{\sqrt{\lambda(M_{2}^2,m_1^2,m_2^2)}}{8 M_{2}^2} \int d\Omega_{2}^*,
\end{eqnarray}
where one first samples the mass $\rm M_{2}$, $\rm M_{3}$\ldots
$M_{n-1}$ in the appropriate limits.

In some topologies symmetric cases of mass generation can appear (as
shown in Figure \ref{f:ttbar}) where the integration sequence is
ambivalent (e.g. in Figure \ref{f:ttbar} the ambivalence is which top
quark invariant mass to generate first\ldots) and after a choice is
made (since one of the two cases in the symmetric pair has to take
precedence) the procedure itself remains not entirely
symmetric. Detailed studies have shown that it proves useful to
include all permutations of such ambiguous sequences into the MC
algorithm in order to 'symmetrise' the solution and thus make it
easier to process by further additions (e.g. adaptive algorithms).

\subsection{The t-type Branching Algorithms}
\begin{figure}[ht]
\begin{center}
     \epsfig{file=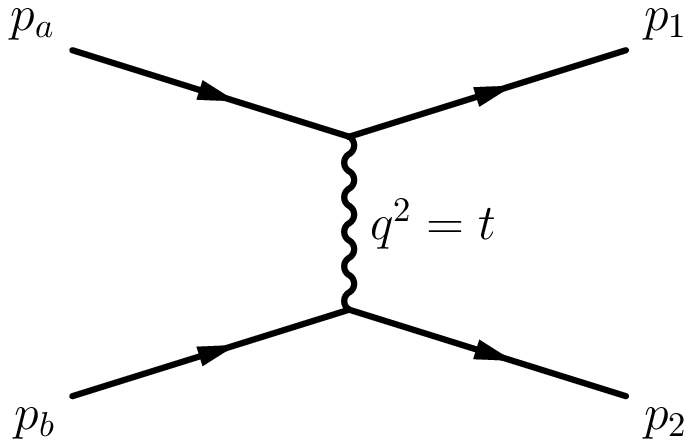,width=3.5cm}
\end{center}
\isucaption{
\small A diagram of a generic $\rm 2 \to 2$ t-channel process.
\label{f:tchan}}
\end{figure}

The t-splits are a specialty of the KB formalism due to the advanced
calculation of the limits on the (massive) t-variable. The formalism can be
introduced by observing that in case of a $\rm p_a + p_b \to p_1 + p_2$
scattering the momentum transfer is characterised by the (Mandelstam) variable
$\rm t = (p_1 - p_a)^2$ (c.f. Fig \ref{f:tchan}). It is thus sensible to replace the $d\Omega_1^* =
d\cos\theta_1^* d\varphi_1^*$ integration in the two body phase space integral of
Eq. \ref{e:phi2s} with integration over the $\rm t$ variable. Writing the definition of
$\rm t$ in the centre-of-mass system one gets:
\begin{eqnarray}
t &=& q^2 = (p_a - p_1)^2 \\ \notag
 &=& m_a^2 + m_1^2 -2E_a^* E_1^* +2p_a^* p_1^* \cos\theta_1^*
\label{e:tdef}
\end{eqnarray}
and hence:
\begin{equation}
 dt = 2 p_a^* p_1^* d\cos\theta^*
\end{equation}
Using the latter substitution together with Eq. \ref{e:cmse},\ref{e:cmsp} and
the analogue for $\rm p_a$:
\begin{equation}
p_a^* = \frac{\sqrt{\lambda(s,m_a^2,m_b^2)}}{2 \sqrt{s}}
\label{e:pa}
\end{equation}
one obtains in place of Eq. \ref{e:phi2s}
\begin{eqnarray}
\Phi_2(s,m_1,m_2)
&=&\frac{p_1^*(s,m_1,m_2)}{4 \sqrt{s}} \int d\Omega_1^* \\  \notag
&=&\frac{1}{8\sqrt{s} p_a^*} \int dt\: d\varphi^*\\ \notag
&=&\frac{1}{4\sqrt{\lambda(s,m_a^2,m_b^2)}} \int\limits_{t^{-}}^{t^{+}} dt \int\limits_0^{2\pi} d\varphi^*
\label{e:phi2t}
\end{eqnarray}
With the integration variable change the integration domain changes from $\rm
[-1,1]$ for $\rm d\cos\theta^*$ to $\rm [t^{-},t^{+}]$ for the $\rm dt$ integration. 
The $\rm t^\pm$ limits are obtained by inserting the $\rm \cos\theta^*$ limits
into Equation \ref{e:tdef}:
\begin{equation}
t^\pm = m_a^2 + m_1^2 -2E_a^* E_1^* \pm 2p_a^* p_1^*
\end{equation}
or in the Lorentz invariant form (c.f. Eq. \ref{e:cmse},\ref{e:cmsp}):
\begin{eqnarray}
t^\pm &=& m_a^2 + m_1^2 - \frac{(s+m_a^2 - m_b^2)(s+m_1^2 -m_2^2)}{2 s} \\ \notag
&\pm & \frac{\sqrt{\lambda(s,m_a^2,m_b^2)\lambda(s,m_1^2,m_2^2)}}{2 s}
\end{eqnarray}

As a step towards generalisation one has to note that the kinematic limits
$t^\pm$ can also be derived from \emph{the basic four-particle kinematic
function G(x,y,z,u,v,w)}\cite{Nyborg65a,KB}, where the function G can be
expressed as a Cayley determinant:
\begin{equation}
G(x,y,z,u,v,w) = -\frac{1}{2} \left|\;
\begin{matrix} 
0 & 1 & 1 & 1 & 1 \\
1 & 0 & v & x & z \\
1 & v & 0 & u & y \\
1 & x & u & 0 & w \\
1 & z & y & w & 0 
\end{matrix}
\; \right|
\end{equation}
The kinematic limits on $\rm t$ are in this case given by the condition 
\begin{equation}
G(s,t,m_2^2,m_a^2,m_b^2,m_1^2) \leq 0,
\end{equation}
it should be noted that the above condition gives either $\rm t^\pm$ limits given a
fixed value of $\rm s$ or equivalently  $\rm s^\pm$ limits given a fixed $\rm t$
value. 

\begin{figure}[ht]
\begin{center}
     \epsfig{file=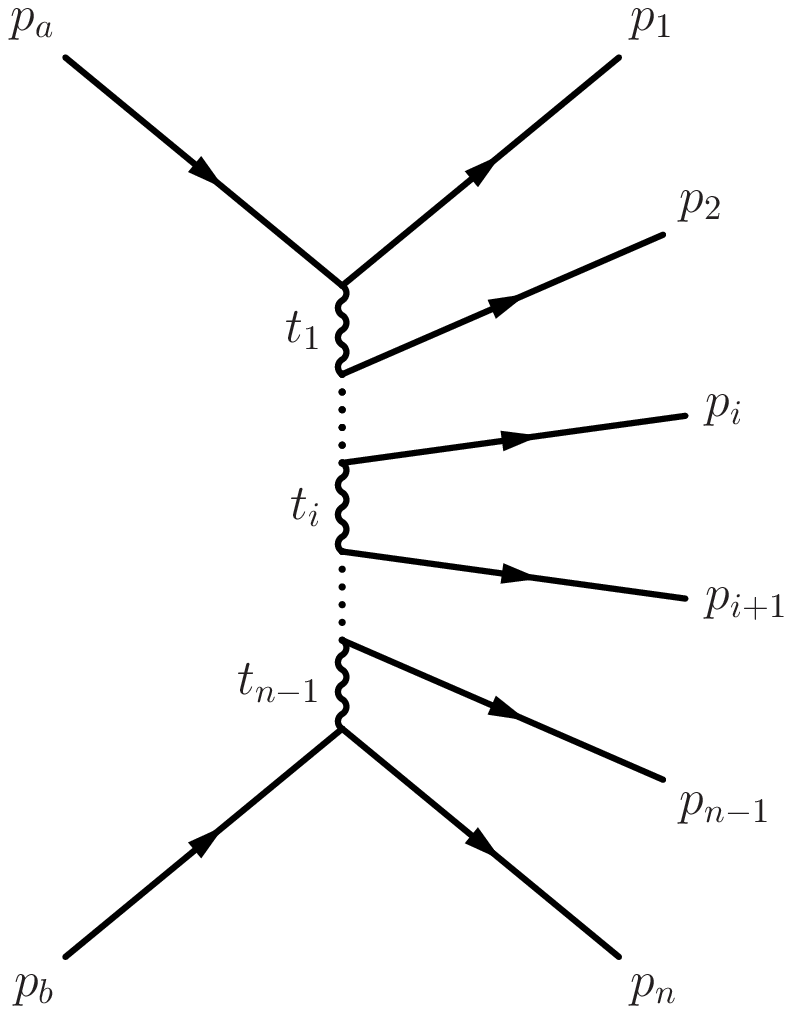,width=6.0cm}\hspace{-1cm}
     \epsfig{file=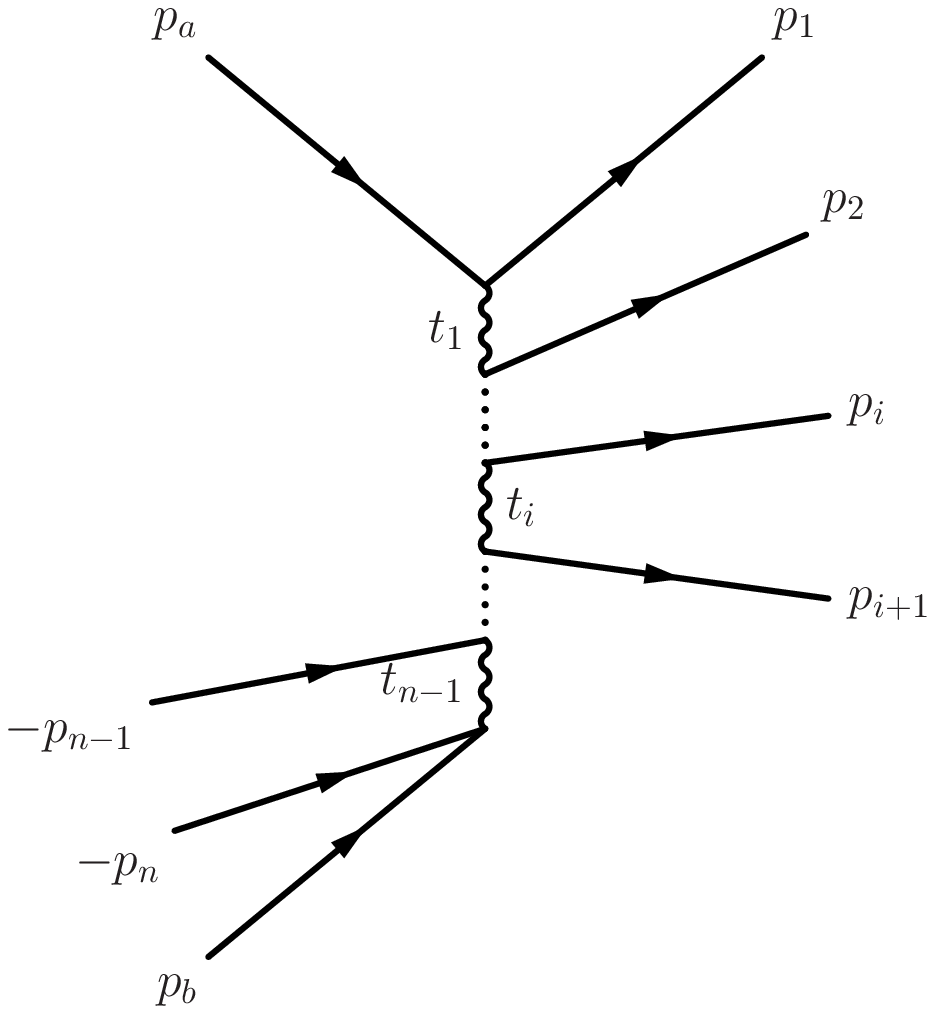,width=6.0cm}\hspace{-1cm}
     \epsfig{file=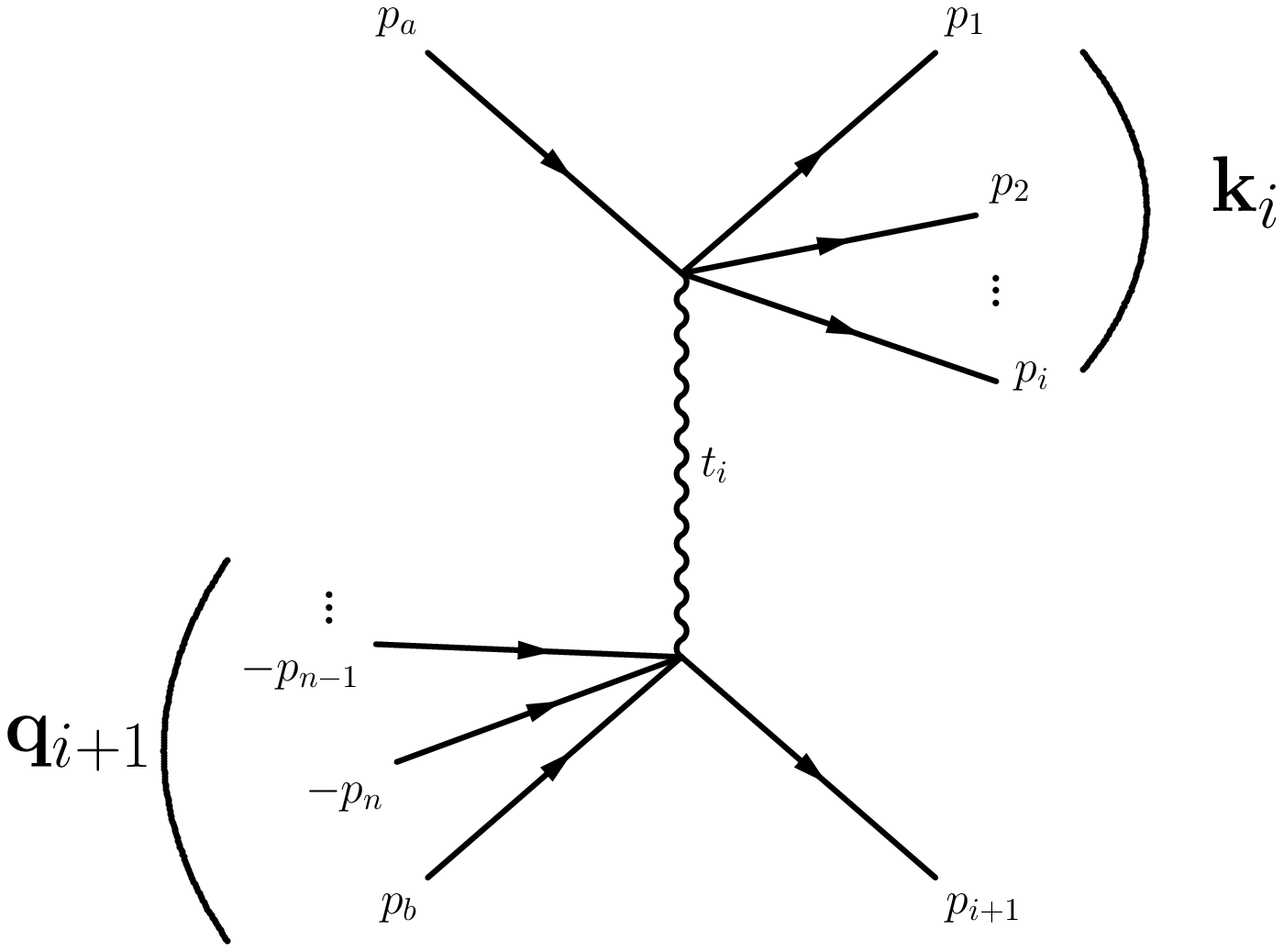,width=6.0cm}
\end{center}
\isucaption{
\small The diagrammatic representation of the method applied in translating the multi-(peri)pheral
splits into a $\rm 2 \to 2$ t-channel configuration.
\label{f:tsplits}} 
\end{figure}

In search of a recursion relation involving t-variables one can note that in
Eq. \ref{e:recur} the angle in $\cos\theta_n^*$ is equivalent to the scattering
angle in the centre-of-mass system of the reaction $\rm p_a + p_b \to k_{n-1} +
p_n$ and thus given by:
\begin{eqnarray}
t_{n-1} &=& (p_a - k_{n-1})^2 \\ \notag
 &=& m_a^2 + M_{n-1}^2 -2E_a^* k_{n-1}^{0*} +2p_a^* k_{n-1}^* \cos\theta_{n-1}^*
\label{e:tndef}
\end{eqnarray}
with the $\rm t_{n-1}^\pm$ limits expressed by:
\begin{equation}
G(M_n^2,t_{n-1},m_n^2,m_a^2,m_b^2,M_{n-1}^2) \leq 0,
\label{e:gtilim}
\end{equation}
and the $p_a^*$ given by Eq. \ref{e:pa}. In order to produce a more general
picture it can further be deduced that the next angle in the recursion
$\theta_{n-1}^*$, is the scattering angle of the subsequent process  
$\rm p_a + (p_b - p_n) \to k_{n-2} + p_{n-1}$ in the centre-of-mass system of
$k_{n-1}$; the $\rm (p_b - p_n)= q_{n-1}$ is in this case considered as a virtual
incoming particle with momentum $\rm q_{n-1}$ (c.f. Figure \ref{f:tsplits}).

\noindent
It immediately follows that for a general process $p_a + q_{i+1} \to k_{i} + p_{i+1}$
with:
\begin{equation}
q_{i} = p_b - \sum_{j=i+1}^{n} p_j = p_a - k_{i}; ~~~~ q_{i}^2 = t_{i}; ~~~
q_n^2 = t_n  = m_b^2
\end{equation}
a general expression for $\rm t_{i}$ becomes in the centre-of-mass frame of $\rm k_{i+1}$:
\begin{eqnarray}
t_{i} &=& (p_a - k_{i})^2 \\ \notag
 &=& m_a^2 + M_{i}^2 -2E_a^{*(i+1)} k_{i}^{0*(i+1)} +2p_a^{*(i+1)} k_{i}^{*(i+1)} \cos\theta_{i}^*
\label{e:tidef}
\end{eqnarray}
where momenta in centre-of-mass frame of $\rm k_{i+1}$, denoted with the 
superscript $\rm *(i+1)$, are given by:
\begin{eqnarray}
k_{i}^{*(i+1)} &=& \frac{\sqrt{\lambda(M_{i+1}^2,M_{i}^2,m_{i+1}^2)}}{2 M_{i+1}}\\
p_a^{*(i+1)}  &=& \frac{\sqrt{\lambda(M_{i+1}^2,m_a^2,t_{i+1})}}{2 M_{i+1}}
\end{eqnarray}
and the corresponding energies  $k_{i}^{0*(i+1)}$  and $\rm E_a^{*(i+1)}$ can
simply be obtained by using  the
analogues of Equations \ref{e:cmse},\ref{e:cmsp} or the usual Einstein
mass-energy relations directly.
The corresponding $\rm t_{i}^\pm$ limits given by:
\begin{equation}
G(M_{i+1}^2,t_{i},m_{i+1}^2,m_a^2,t_{i+1},M_{i}^2) \leq 0,
\end{equation}
and the recursion relation of Eq. \ref{e:recur} becomes:
\begin{eqnarray}
\Phi_n(M_n^2,m_1,m_2,\ldots,m_n) &=& \label{e:recurt} \\
&=& \int\limits_{(\sum_{i=1}^{n-1}m_i)^2}^{(M_n - m_n)^2} 
\frac{dM_{n-1}^2 }{4\sqrt{\lambda(M_n^2,m_{a}^2,t_n)}} \int\limits_0^{2\pi} d\varphi_n^*
\int\limits_{t_{n-1}^{-}}^{t_{n-1}^{+}} dt_{n-1}\: 
 \Phi_{n-1}(M_{n-1}^2,m_1,m_2,\ldots,m_{n-1}), \notag
\end{eqnarray}

As already argued the resulting set of $\rm (s_i=M_i^2,t_i)$ can again be
sampled in any direction with respect to the cascade by applying the appropriate
change in the integration limits (c.f. Eq. \ref{e:recur} and \ref{e:revlim}). 
The recommended approach (i.e. the introduced modification of the
algorithm) is again to first sample the invariant masses in the reverse cascade
direction (i.e. in the sequence $\rm M_2,M_3,\ldots ,M_n$) and then the $t_i$
values within the limits calculated from Eq. \ref{e:gtilim} down the cascade
(i.e. in the order of $\rm t_{n-1},t_{n-2},\ldots,t_1$).

To sum up, it has been shown that using the Kajantie--Byckling formalism 
the phase space for any topology can be split in a set
of s-type and t-type $\rm 2 \to 2$ branching steps (modules) 
given by recursive formulae of Equations \ref{e:revlim} and \ref{e:recurt}.
\clearpage

\section{Propagator Sampling\label{s:prop}}

A well known theoretical issue is that one can expect the most prominent peaks
in the differential cross-section of a specific process in the phase space
regions of high propagator values in the corresponding probability
density. Consequently, in the scope of complementing the modular structure of
the derived Kajantie-Byckling based phase space sampling, new approaches were also
developed concerning the numerical sampling methods of the relevant kinematic
quantities.

In order to get small variance in the Monte Carlo procedure one would thus like
to include the appropriate peaking dependence of the relevant momentum transfers
$\rm q^2$ in the importance sampling function. It however turns out that since
the momenta transfers $\rm q$ participate also in the propagator numerators
(typically in $\rm p_\mu q^\mu/q^2$) and since in process of interest one mostly
finds several Feynman diagrams contributing to the final probability density,
thus causing interferences, it is very difficult or even impossible to estimate
the exact power of momenta transfers in the sampling functions for different
propagator peaks. In other words, the probability density dependence on the momentum
transfer $\rm q^2$ can in general be approximated with the dependence $\rm
1/(q^2)^\nu$ where the best value of $\rm \nu$ must be determined separately (on
a process by process basis).

In view of the latter, general formulae have been developed for sampling the
$x^{-\nu}$ shape \cite{fermisv,sampler}: Given a pseudo random number $r \in [0,1]$ and limits 
$x \in [x_{-},x_{+}]$ the value x distributed as $x^{-\nu}$ is obtained from the
formulae in Eq. \ref{e:usampling} as:
\begin{eqnarray}
x &=& \left[ x_{-}^{-\nu+1}\cdot(1-r) + x_{-}^{-\nu+1}\cdot r
\right]^{-\frac{1}{\nu+1}}; ~~~~ \nu \neq 1;\\ x &=&
\frac{x_{+}^r}{x_{-}^{r-1}}; ~~~ \nu = 1.
\label{e:xnuni}
\end{eqnarray}  

\noindent
Using the analogous (unitary) approach a recipe for resonant (Breit-Wigner)
propagator contributions of the type:
\begin{equation}
BW(s)=\frac{1}{(s-M^2)^2+M^2 \Gamma^2}
\end{equation}
with $\rm s \in [s_{-},s_{+}]$ and a pseudo random number $r \in [0,1]$ is
available by the prescription:
\begin{eqnarray}
s &=& M^2 + M\Gamma \cdot \tan \left[ (u_{+} - u_{-}) \cdot r + u_{-} \right]\\
u^\pm &=& {\rm atan}\left( \frac{s^\pm - M^2}{M\Gamma} \right)
\label{e:bwuni}
\end{eqnarray}

Following similar arguments as for the non-resonant propagators one can surmise
that the best sampling function for resonant propagators could in general be a 
Breit-Wigner shape modified by a factor $\rm s^\nu, \nu \in [0,1]$. In \cite{acermc} it was found that
a shape:
\begin{equation}
BW(s)=\frac{s}{(s-M^2)^2+M^2 \Gamma^2}
\end{equation}
works quite well for a set of processes and a corresponding sampling recipe was developed.
In addition, studies in \cite{lichard} show that a resonant $\rm \sqrt{s} \times$ Breit-Wigner shape:
\begin{equation}
BW(s)=\frac{\sqrt{s}}{(s-M^2)^2+M^2 \Gamma^2}
\end{equation}
should be expected in a range of decay processes. Detailed studies have shown
that it is in general better to introduce a $\rm s^\nu, \nu \in [0,1]$
dependence even if it over-compensates the high mass tails of the corresponding
differential cross-section distribution since this provides an overall reduction
of the maximal weight fluctuations in the Monte--Carlo event generation procedure. 

\subsection{The Inclusion of Mass Effects in Propagator Sampling}

Studies have shown that the $x^{-\nu}$ approximation works quite well for
t-channel type propagators since the phase space suppression factor
$\sqrt{\lambda}$ participates in the denominator, as shown in
Eq. \ref{e:recurt}, and thus contributes only to the $x^{-\nu}$
slope. Contrariwise, while the $x^{-\nu}$ approximation still works reasonably
well when describing the s-channel type propagators involving particles with
high virtuality and/or decay products with low masses, it can be shown that this
is not necessarily the case in the LHC environment, where the presence of
massive decay products can significantly affect the invariant mass
distributions.  As it can be seen in Figure \ref{f:acermc} the shape of the
propagator dependence can be strongly suppressed by the phase space $\rm
\sqrt{\lambda}$ (boost) factor at low values; thus the sampling function
approximation for non-resonant propagators could be approximated with something
like:
\begin{equation}
f_{\rm NR}(s) =  \frac{\sqrt{\lambda(s,m_a^2,m_b^2)}}{s} \cdot \frac{1}{s^\nu} = \frac{\sqrt{\lambda(s,m_a^2,m_b^2)}}{s^{\nu+1}}
\label{e:lamxnu}
\end{equation}

and similarly
\begin{equation}
f_{\rm R}(s) = \frac{\sqrt{\lambda(s,m_a^2,m_b^2)}}{s} \cdot \frac{\sqrt{s}}{(s-M^2)^2+M^2 \Gamma^2} = \frac{\sqrt{\lambda(s,m_a^2,m_b^2)}}{\sqrt{s}\cdot((s-M^2)^2+M^2 \Gamma^2)}
\label{e:lamsbw}
\end{equation}

for resonant propagators. 

As it turns out the two functions cannot be sampled by the well known
unitary algorithms (i.e. the biggest collection of recipes \cite{sampler}
yielded no results); already the integral values of the functions yield
complicated expressions which cannot be easily calculated, let alone 
inverted analytically. The solution was to code numerical algorithms
to calculate the integrals (i.e. cumulants) explicitly.

After the integrals are calculated, their inverse and the subsequent
sampling value can again be obtained numerically. Namely, resorting to
the original definition of the unitarity sampling recipe in
Eq. \ref{e:usampling}, by replacing the normalised $\rm g_i(x)$ with:
\begin{equation}
 g_i(x) \to \frac{f(x)}{\int\limits_{x_{-}}^{x_{+}} f(x)\, dx},
\end{equation}
which in turn gives:
\begin{equation}
\int\limits_{x_{-}}^{x} f(x)\; dx = r \cdot \int\limits_{x_{-}}^{x_{+}} f(x)\; dx,
\label{e:sampling}
\end{equation}
where $\rm f(x)$ is the non-negative function one wants to sample from, 
$\rm [x_{-},x_{+}]$ is the range of values of the parameter $\rm x$ we want to sample 
and $\rm r$ a pseudo random number $r \in [0,1]$. As already stated
(c.f. Eq.\ref{e:usampling}), in the case when the integral of the function $\rm f(x)$ 
is an analytic function, $F(x) = \int_{x_{-}}^x f(x)\, dx$, and has a known inverse $\rm F^{-1}(x)$ one 
can construct explicit unitary prescriptions by:
\begin{equation}
x = F^{-1}\left( r \cdot [ F(x_{+})-F(x_{-}) ] + F(x_{-}) \right)
\end{equation}
as given for two particular cases in Eq. \ref{e:xnuni},\ref{e:bwuni}.

In the cases the integral can not be inverted, the prescription of the
Eq. \ref{e:sampling} can directly be transformed into a zero-finding request;
thus, since both the integral and the first derivative (i.e. the sampling
function and its cumulant) are known, the Newton-Rhapson method is chosen as the
optimal one for root finding:

\begin{eqnarray}
g(x) &=& \left\{ \int\limits_{x_{-}}^{x} f(x) dx - r \cdot \int\limits_{x_{-}}^{x_{+}} f(x) dx \right\} = 0\\
g'(x)  &=& \frac{d}{dx} \left\{ \int\limits_{x_{-}}^{x} f(x) dx - r \cdot \int\limits_{x_{-}}^{x_{+}} f(x) dx \right\} =  f(x)
\end{eqnarray}

With a sensible choice of starting points the procedure generally takes
on the order of ten cycles until finding the root with adequate numerical
precision. The overall generation speed is still deemed quite
reasonable. 

\noindent
The integration of the phase-space suppressed resonant propagator of
Eq. \ref{e:lamsbw} yields a rather non-trivial expression:
\begin{eqnarray}
\int\limits_{(m_a+m_b)^2}^{s}f_{\rm R}(s)~ ds &=& \int\limits_{(m_a+m_b)^2}^{s} \frac{\sqrt{\lambda(s,m_a^2,m_b^2)}\, ds}{\sqrt{s}\cdot((s-M^2)^2+M^2 \Gamma^2)}\\  \notag
&=& \int\limits_{a}^{s}\frac{\sqrt{(s-a)(s-b)}~ ds}{\sqrt{s}\cdot((s-M^2)^2+M^2 \Gamma^2)}\\ \notag
&=& \frac{1}{{\sqrt{-b}}\,\Gamma\,M^2} \times  \frac{-2\,i \,a\,b\,\Gamma\,}{( \Gamma^2 + M^2 ) }\\ \notag  &\times& 
\biggl\{  
           \Mfunction{F}\left[ i \,{\rm arcsinh}(\frac{{\sqrt{-b}}}{{\sqrt{a}}}),\frac{a}{b} \right]
-\Mfunction{F}\left[ i \, {\rm arcsinh}(\frac{{\sqrt{-b}}}{{\sqrt{s}}}),\frac{a}{b}\right] \\ \notag
           &+& ( i \,\Gamma + M ) \, 
           ( a + i \,( \Gamma + i \,M ) \,M ) \,
           ( b + i \,( \Gamma + i \,M ) \,M ) \, 
           \Mfunction{\Pi}\left[ \frac{M\,( -i \,\Gamma + M ) }{b}, i \,{\rm arcsinh}(\frac{{\sqrt{-b}}}{{\sqrt{a}}}),
            \frac{a}{b} \right] \\  \notag
           &+& ( \Gamma + i \,M ) \,
           ( b + ( -i \,\Gamma - M ) \,M ) \,
           ( i \,a + ( \Gamma - i \,M ) \,M ) \,
           \Mfunction{\Pi}\left[ \frac{M\,( i \,\Gamma + M ) }{b},
            i \,{\rm arcsinh}(\frac{{\sqrt{-b}}}{{\sqrt{a}}}),
            \frac{a}{b}\right]  \\  \notag
           &-& ( i \,\Gamma + M ) \,
           ( a + i \,( \Gamma + i \,M ) \,M ) \,
           ( b + i \,( \Gamma + i \,M ) \,M ) \,
           \Mfunction{\Pi}\left[ \frac{M\,( -i \,\Gamma + M ) }{b},
            i \,{\rm arcsinh}(\frac{{\sqrt{-b}}}{{\sqrt{s}}}),\frac{a}{b}\right]\\  \notag
           &-& ( \Gamma + i \,M ) \,
           ( b + ( -i \,\Gamma - M ) \,M ) \,
           ( i \,a + ( \Gamma - i \,M ) \,M ) \,
           \Mfunction{\Pi}\left[ \frac{M\,( i \,\Gamma + M ) }{b},
            i \,{\rm arcsinh}(\frac{{\sqrt{-b}}}{{\sqrt{s}}}),
            \frac{a}{b}\right]   \biggr\}
\end{eqnarray}
where the variables $a,b$ stand for $\rm a=(m_a+m_b)^2$ and $\rm b=(m_a-m_b)^2$
and the functions $\rm \Mfunction{F}[\varphi,k]$ and $\rm
\Mfunction{\Pi}[\varphi,k,n]$ are the Legendre's incomplete elliptic integrals
of the second and third kind with complex arguments. In order to perform the
calculations the latter functions had to be coded from scratch since they were
not found in any (publicly available) computer libraries or code repositories.
The prescriptions for calculating them were found in \cite{carlson}; the results
were checked against the values given by {\tt Mathematica$^{\rm TM}$}.

\noindent In the special case $\rm m_a = m_b$ the above expression simplifies into:
\begin{eqnarray}
\int\limits_{(2 m_a)^2}^{s}f_{\rm R}(s)~ ds &=& \int\limits_{(2 m_a)^2}^{s} \frac{\sqrt{\lambda(s,m_a^2,m_a^2)}\, ds}{\sqrt{s}\cdot((s-M^2)^2+M^2 \Gamma^2)}\\ \notag
&=& \int\limits_{a}^{s}\frac{\sqrt{(s-a)}~ ds}{\cdot((s-M^2)^2+M^2 \Gamma^2)}\\ \notag
&=&  \frac{1}{\Gamma\,M\,
     {\sqrt{a + \left( -i \,\Gamma - M \right) \,M}}}\\ \notag
&\times& 
\biggl\{
\left( i \,a + \left( \Gamma - i \,M \right) \,M \right) \,
      \arctan (\frac{{\sqrt{-a + z}}}
        {{\sqrt{a + \left( -i \,\Gamma - M \right) \,M}}}) \\ \notag
&-& 
     i \,{\sqrt{a + \left( -i \,\Gamma - M \right) \,M}}\,
      {\sqrt{a + i \,\left( \Gamma + i \,M \right) \,M}}\,
      \arctan (\frac{{\sqrt{-a + z}}}
        {{\sqrt{a + i \,\left( \Gamma + i \,M \right) \,M}}})  \biggr\}
\end{eqnarray}

\noindent The result of integrating the phase-space suppressed non-resonant
propagator (Eq. \ref{e:lamxnu}) yields a similarly non-trivial result:

\begin{eqnarray}
\int\limits_{(m_a+m_b)^2}^{s} f_{\rm NR}(s)~ds &=& \int\limits_{(m_a+m_b)^2}^{s} \frac{\sqrt{\lambda(s,m_a^2,m_b^2)}~ds}{s^{\nu+1}}\\
&=& \notag
  \frac{1}{2\,{\sqrt{1 - \frac{s}{a}}}\,\nu}  \biggl\{
\frac{-2\,{\sqrt{\left( a - s \right) \,\left( b - s \right) }}\,
         \Mfunction{F_1}\left[ -\nu,-\left( \frac{1}{2} \right) ,
         -\left( \frac{1}{2} \right) ,1 -
         \nu,\frac{s}{a},\frac{s}{b}\right]}{s^\nu\, {\sqrt{1 -
         \frac{s}{b}}}} \\  \notag &+& \frac{{\sqrt{\pi }}\,{\sqrt{\left( -a +
         b \right) \, \left( a - s \right)
         }}\,\Mfunction{\Gamma}\left[1 - \nu\right]\,
         \Mfunction{F}\left[-\nu,-\left( \frac{1}{2} \right) ,
         \frac{3}{2} - \nu,\frac{a}{b}\right]}{a^\nu\,{\sqrt{1 -
         \frac{a}{b}}}\, \Mfunction{\Gamma}\left[\frac{3}{2} -
         \nu \right]} \biggr\} 
\end{eqnarray}

where the function $\rm F[\alpha,\beta,\gamma,x]$ is the Gauss Hypergeometric
function and the $\rm F_1[\alpha,\beta,\beta',\gamma,x,y]$ is the two-parameter
(Appell) Hypergeometric function \cite{appell}. Both functions can be calculated
by using the prescriptions in \cite{appell}; it however turns out that the
calculation of the $\rm F_1[\alpha,\beta,\beta',\gamma,x,y]$ to a certain (high)
precision is almost two times slower than the explicit numerical calculation of
the integral to the same precision. Subsequently the numerical evaluation of the
Gauss Hypergeometric function $\rm F[\alpha,\beta,\gamma,x]$ was retained since
it participates in the $\rm m_a = m_b$ simplification and the calculation of the
integral was done by using a 50-point Gauss-Legendre quadrature with $\sqrt{s}$
weight function; the weights were calculated by \cite{quad}. The implementation of the
(Appell) Hypergeometric function calculation was used as a cross-check to confirm the
correct implementation and precision of the numerical method.

\noindent As already mentioned, the above integral again simplifies for $\rm m_a = m_b$:
\begin{eqnarray}
\int\limits_{(2 m_a)^2}^{s} f_{\rm NR}(s)~ds &=& \int\limits_{(2 m_a)^2}^{s} \frac{\sqrt{\lambda(s,m_a^2,m_b^2)}~ds}{s^{\nu+1}} \\ \notag
&=& \int\limits_{a}^{s} \frac{\sqrt{(s-a)}~ds}{s^{\nu+\frac{1}{2}}} \\ \notag
&=& \frac{2}{3}\,a^{1-\nu}\,s^{\frac{3}{2}}\,\Mfunction{F}\left[ \frac{3}{2},\nu+\frac{1}{2},
      \frac{5}{2},-s \right], 
\end{eqnarray}
and the Gauss Hypergeometric function $\rm F[\alpha,\beta,\gamma,x]$ is in this case
calculated by the methods described in  \cite{appell} with some improvements analogous to
the ones described e.g. in \cite{numrec}. 
\clearpage

\section{Application of the Method \label{s:num2}}

The phase space 'modularisation' described in this paper has successfully been
applied for phase-space generation in the {\tt AcerMC 2.0} Monte-Carlo
generator\cite{acermc2}. The program uses the multi-channel phase space generation where each
channel corresponds to an expected phase space topology as derived from the
participating Feynman diagrams. In the {\tt AcerMC 2.0} this information was
obtained from the modified {\tt MadGraph}\cite{madgraph} program which also
supplied the probability amplitudes for the implemented processes. Each channel
topology was in turn constructed from the t-type and s-type modules and sampling
functions described in this paper together with some additional importance
sampling techniques for space angles and rapidity distributions described in
detail elsewhere \cite{pythia,fermisv,excal,acermc}. The unknown slope
parameters (denoted $\nu$ in the text, c.f. Eq. \ref{e:lamxnu}) of the invariant
mass sampling functions for non-resonant propagators were obtained by short training
runs of the program on a process by process basis. 

\begin{figure}[ht]
\begin{center}
     \epsfig{file=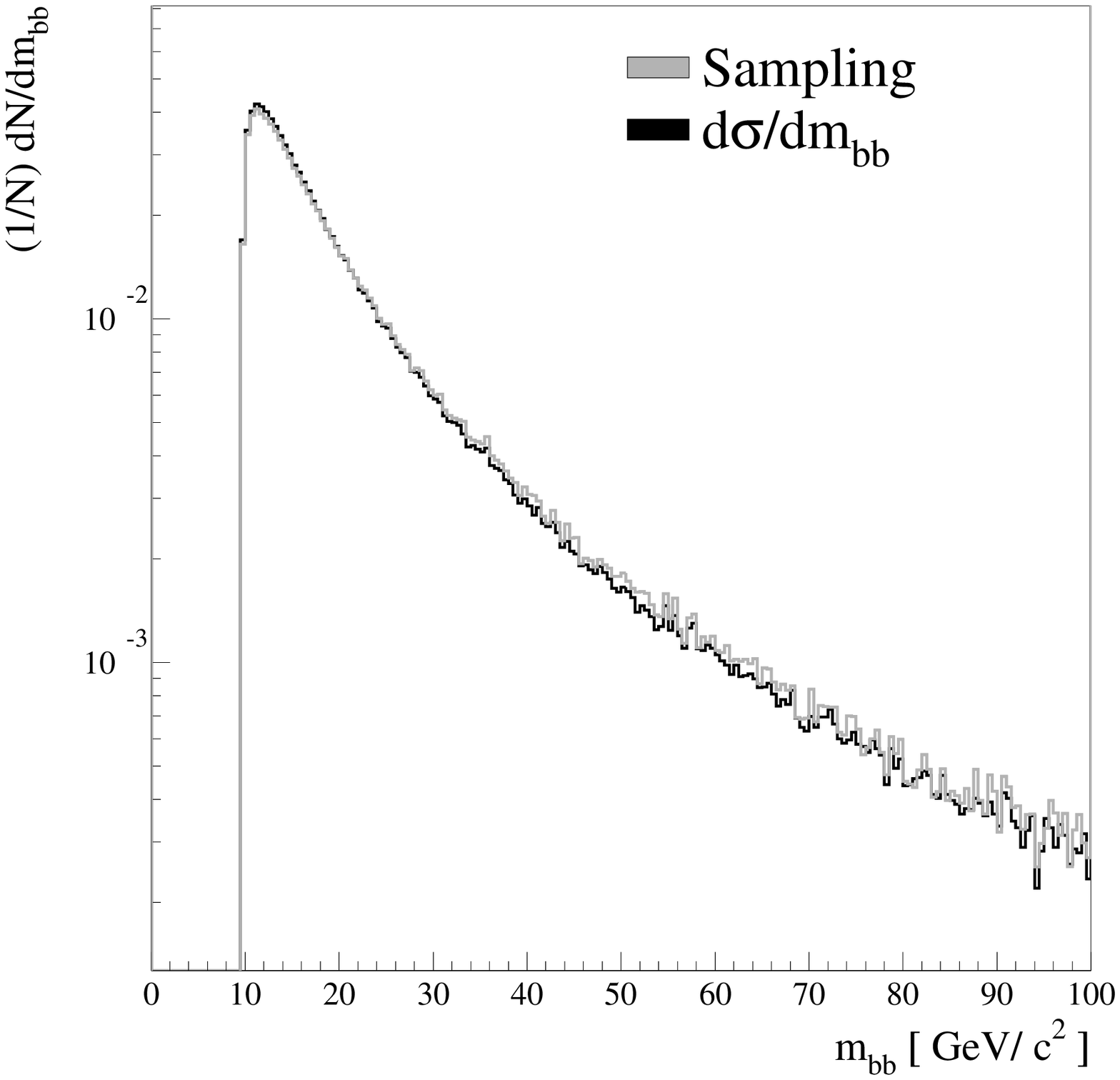,width=5.3cm}\hspace{-0.3cm}
     \epsfig{file=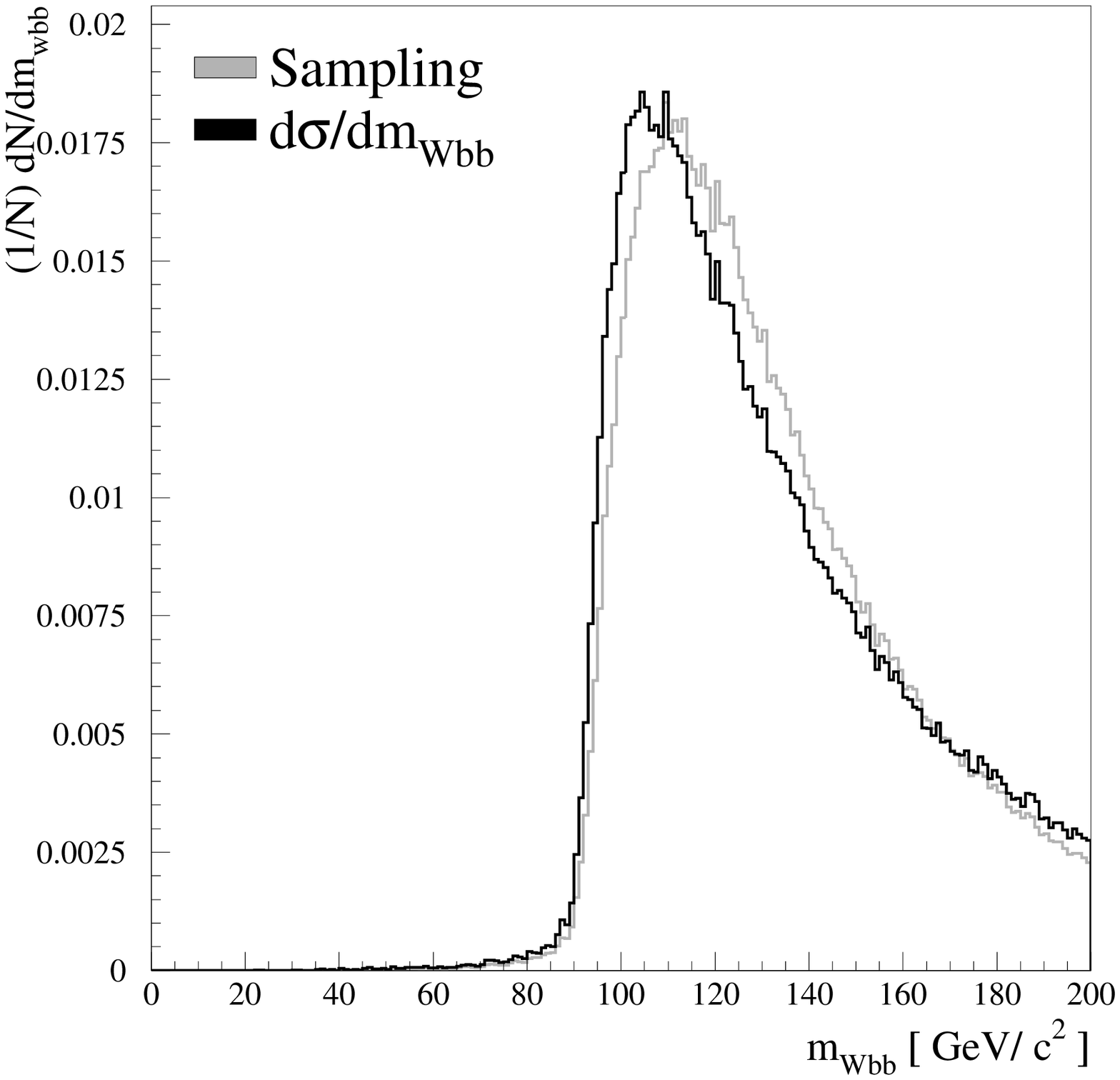,width=5.3cm}\hspace{-0.3cm}
     \epsfig{file=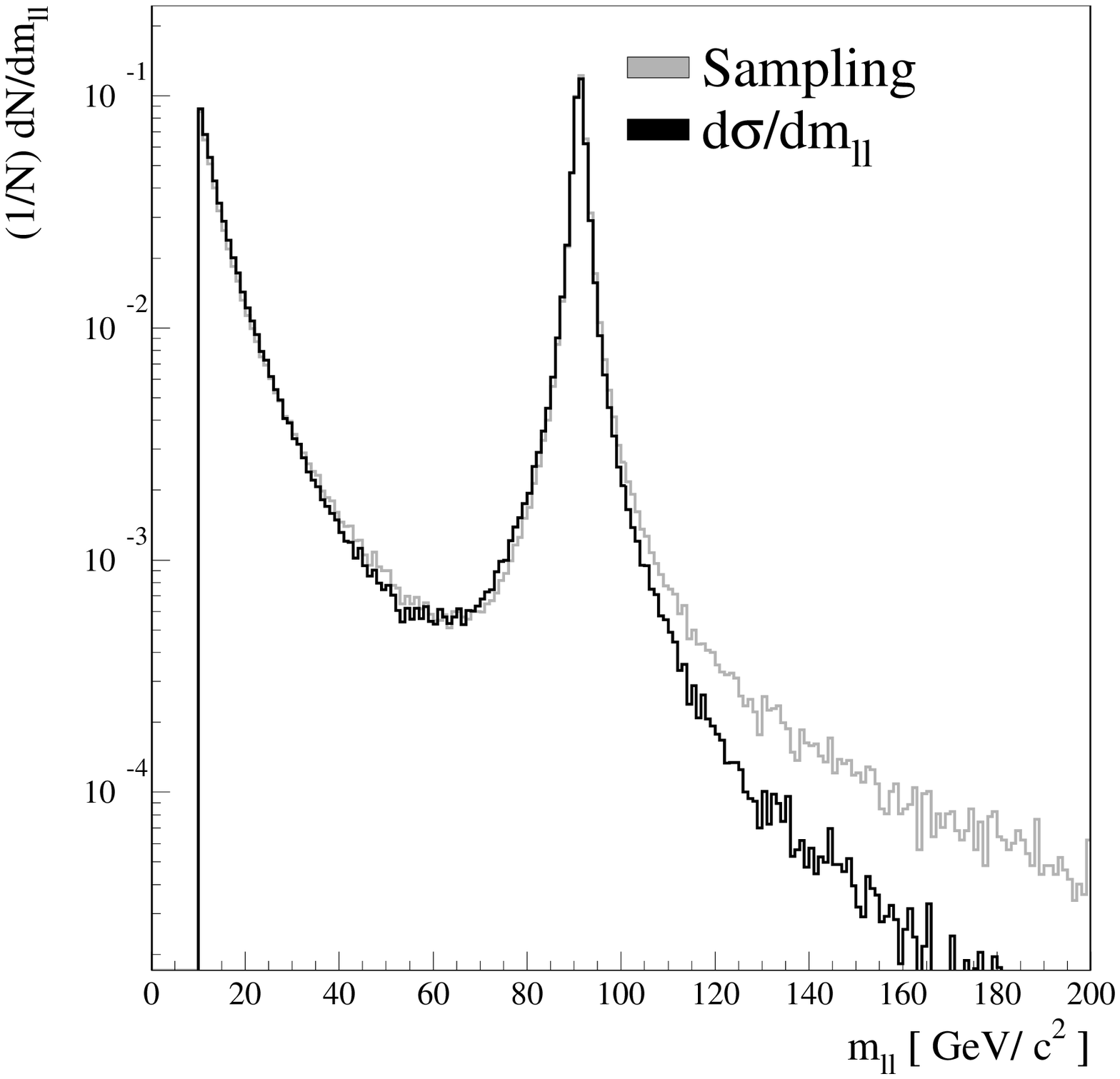,width=5.3cm}\vspace{-0.5cm}
\end{center}
\isucaption{
\small A few representative invariant mass distribution comparisons between the
(normalised) sampling functions and the normalised differential cross-section as obtained
with {\tt AcerMC 2.0} Monte--Carlo generator.  Left: The invariant mass of the
$\rm b \bar{b}$ pair in the process $\rm u \bar{d} \to W^+ g^* \to l^+ \nu_l b
\bar{b}$.  Center: The invariant mass of the $\rm W b \bar{b}$ system
(equivalently the hard centre-of-mass energy $\rm \sqrt{\hat{s}}$) for the same
process.  Right: The invariant mass of the $\rm \ell \bar{\ell}$ pair in the
process $g g \to Z^0/\gamma^* b \bar{b} \to \ell \bar{\ell} b \bar{b}$. All the
distributions were obtained using the prescriptions of this paper without the
adaptive algorithms also used in the {\tt AcerMC 2.0} Monte--Carlo generator. As
one can see the approximations used seem to work quite well.
\label{f:acermc}} 
\end{figure}

As a further step the multi-channel self-optimisation procedure was implemented
in order to minimise the variance of the event weights further
\cite{kleiss}.  A few representative invariant mass distribution comparisons
between the implemented sampling functions and the actual differential
distributions are shown in Figure \ref{f:acermc}.

As a further estimate of the success of the methods a comparison of the 
variance in the differential cross-section determination are presented in Table
\ref{t:comps} for {\tt AcerMC 2.0} \cite{acermc2} and an earlier version, {\tt
AcerMC 1.4}  \cite{acermc},
which uses the standard phase space sampling techniques
\cite{fermisv,excal,nexcal,bhlumi,acermc}. The comparison is done for a few
representative processes. It has to be stressed that in {\tt AcerMC
1.x} versions the phase space was constructed by hand on a process by
process basis meticulously tuned while in the new {\tt AcerMC 2.0} the
'automated' approach sufficed. A further improvement was the
inclusion of the ac-VEGAS algorithm \cite{acermc} which reduces the
maximal event weights in order to improve the unweighing
efficiencies; the comparison between the unweighing efficiencies
reached in the standard and new approaches ({\tt AcerMC} versions {\tt
1.4} and {\tt 2.0}) is also given in Table \ref{t:comps}. The event
weight variance $\rm V_\sigma$ presented in Table \ref{t:comps} is 
for $\rm N$ measurements of the weights $\rm w_i$ defined as:
\begin{equation}
V_\sigma = \frac{\frac{\sum_{i=1}^{N} w_i^2}{N} - \left( \frac{\sum_{i=1}^{N} w_i}{N} \right)^2}{N-1}.
\end{equation}
Likewise, the customary definition is used for the unweighing efficiency estimate:
\begin{equation}
\epsilon = \frac{\frac{\sum_{i=1}^{N} w_i}{N}}{w_\text{max}}
\end{equation}
where the $\rm w_\text{max}$ is the maximal event weight obtained in $\rm N$ trials.
It also might be relevant to stress that since the average weight is the total cross-section estimate for the
considered process $\rm \sigma \simeq \sum_{i=1}^{N} w_i/N$ the only quantity that is allowed to 
change in order to improve the unweighing efficiency is the maximal weight $\rm w_\text{max}$.

\begin{table}[htb]
\small
\newcommand{\lstrut}{{$\strut\atop\strut$}}
  \isucaption {\small The process cross-section variances with their
  uncertainties and unweighing efficiencies obtained for a few
  implemented processes basing on the old and new phase space sampling
  techniques as implemented in {\tt AcerMC 1.4} and {\tt AcerMC 2.0}
  respectively. The variances are given for a sequence of $\rm 10^5$
  weighted events (i.e. algorithm iterations) obtained by using the
  procedure described in the text. The unweighing efficiencies were
  estimated from samples containing $\rm \sim 10^6$ weighted
  events. \label{t:comps}}
\vspace{2mm}  
\begin{center}
\begin{tabular}{lcccc}
\hline\noalign{\smallskip}
Process &  {\tt AcerMC 2.0} $\rm V_\sigma$  $\rm [ pb^2 ]$ &  {\tt AcerMC 1.4} $\rm V_\sigma$ $\rm [ pb^2 ]$ & {\tt AcerMC 2.0} $\rm \epsilon  $ &
{\tt AcerMC 1.4} $\rm \epsilon$ \\
\noalign{\smallskip}\hline\noalign{\smallskip}
$gg \to Z/(\to \ell \ell)  b \bar b $      & $0.129\cdot 10^{-2}  \pm 0.52\cdot 10^{-5} $ &  $0.159\cdot 10^{-2} \pm 0.61\cdot 10^{-5}$ & 37\% & 33\% \\
\noalign{\smallskip}\hline\noalign{\smallskip}
$q \bar q  \to W(\to \ell \nu)  b \bar b $ & $0.390\cdot 10^{-2}  \pm 0.15\cdot 10^{-4}$  &  $ 0.533\cdot 10^{-2} \pm 0.18\cdot 10^{-4}$ & 35\% & 33\% \\
\noalign{\smallskip}\hline\noalign{\smallskip}
$gg  \to   t \bar t b \bar b $             & $0.522\cdot 10^{-4}  \pm 0.19\cdot 10^{-6} $  & $0.972\cdot 10^{-4} \pm 0.44\cdot 10^{-6} $ & 36\% & 20\% \\
\noalign{\smallskip}\hline
\end{tabular}
\end{center} 
\end{table}

\begin{table}[htb]
\small
\newcommand{\lstrut}{{$\strut\atop\strut$}}
  \isucaption {\small The process cross-sections and variances with
  their uncertainties and unweighing efficiencies as obtained for two
  sample $\rm 2 \to 6$ processes implemented in {\tt AcerMC 2.0}
  Monte--Carlo generator.  The results show
  that the sampling procedure scales quite efficiently with the
  increase in the number of Feynman diagrams and sampling channels. Detailed
  studies show that with inclusion of the new propagator sampling the
  angular distributions become the restraining factor. The
  cross-sections and variances are given for a sequence of $\rm 2
  \cdot 10^5$ weighted events (i.e. algorithm iterations) obtained by
  using the procedure described in the text. The unweighing
  efficiencies were estimated from samples containing $\rm \sim 10^6$
  weighted events. \label{t:comps2}}
\vspace{2mm}  
\begin{center}
\begin{tabular}{lcccc}
\hline\noalign{\smallskip}
{\tt AcerMC 2.0} Process & & $\rm \sigma$ $\rm [ pb]$ & $\rm V_\sigma$  $\rm [ pb^2 ]$ &   $\rm \epsilon  $ \\
\noalign{\smallskip}\hline\noalign{\smallskip}
 $gg \to t\bar{t} \to b \bar{b} W^+ W^- \to b \bar{b} \ell \bar{\nu}_\ell \bar{\ell} \nu_\ell$ &(3 Feyn. diag./2 sampl. chan.)
  & $ 4.49 $ & $ 0.80 \cdot 10^{-4}  \pm 0.39 \cdot 10^{-6} $ & 14\% \\[8pt]
 $gg \to b \bar{b} W^+ W^- \to b \bar{b} \ell \bar{\nu}_\ell \bar{\ell} \nu_\ell$ &(31 Feyn. diag./13 sampl. chan.)
  & $4.77$  & $ 0.77 \cdot 10^{-4}  \pm 0.29 \cdot 10^{-5} $ & 17\% \\
\noalign{\smallskip}\hline
\end{tabular}
\end{center} 
\end{table}

\begin{figure}[ht]
\begin{center}
     \epsfig{file=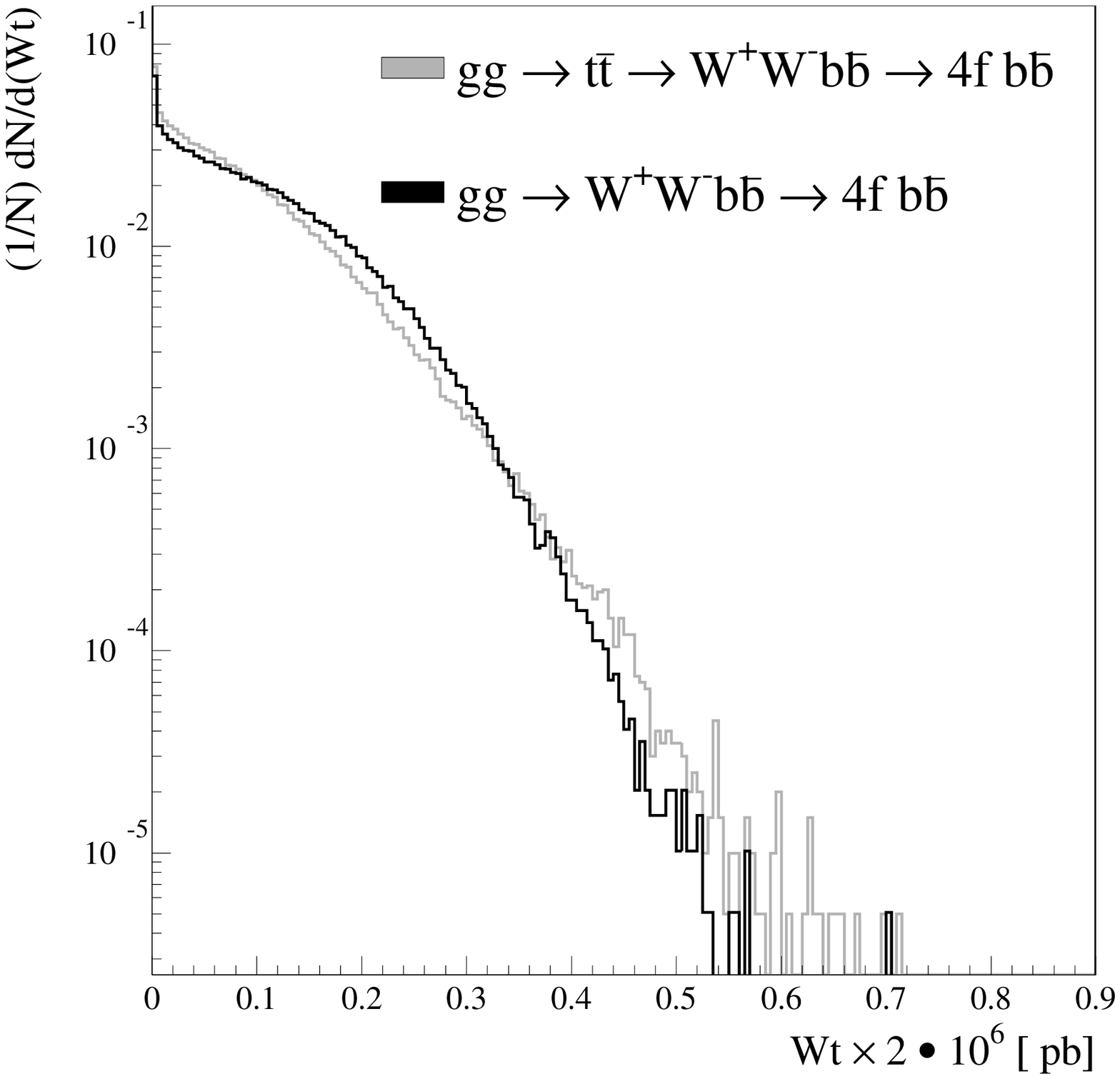,width=6cm}
\end{center}
\isucaption{
\small The weight distribution of the sampled events for the   
$gg \to t\bar{t} \to b \bar{b} W^+ W^- \to b \bar{b} \ell
\bar{\nu}_\ell \bar{\ell} \nu_\ell$ (light gray histogram) and $gg \to
 b \bar{b} W^+ W^- \to b \bar{b} \ell \bar{\nu}_\ell
\bar{\ell} \nu_\ell$ (black histogram) processes as obtained with {\tt
AcerMC 2.0} Monte--Carlo generator. One can observe the well defined
weight range for the two processes; as it turns out the weight
distribution is even marginally better for the (more complex) second
process, possibly because the higher number of sampling channels
manage to cover the event topologies in phase space to a better
extent. \label{f:acermc2}}
\end{figure}

The improved and automated phase space handling provided the means
to include the $\rm 2 \to 6$ processes like e.g. $gg \to t\bar{t} \to b \bar{b}
W^+ W^- \to b \bar{b} \ell \bar{\nu}_\ell q_1 \bar{q}_2$
(c.f. Fig. \ref{f:ttbar}) which would with the very complicated phase space
topologies prove to be too much work to be handled manually. The studies show
that the overall unweighing efficiency which can be reached in the $\rm 2 \to 6$
processes by using the recommended phase space structuring is on the order of
$\rm 10$ percent. As an example, process cross-sections and variances with their
uncertainties and unweighing efficiencies as obtained for two sample $\rm 2 \to
6$ processes implemented in {\tt AcerMC 2.0} Monte--Carlo generator are
presented in Table
\ref{t:comps2} and the corresponding weight distributions are shown in Figure \ref{f:acermc2}.
The two processes share the same initial and final state but include different
subsets of Feynman diagrams; the $gg \to t\bar{t}
\to b \bar{b} W^+ W^- \to b \bar{b} \ell \bar{\nu}_\ell
\bar{\ell} \nu_\ell$ process includes only the three diagrams
containing the $ t \bar{t}$ intermediate state while in the second process $gg
\to b \bar{b} W^+ W^- \to b \bar{b} \ell \bar{\nu}_\ell
\bar{\ell} \nu_\ell$ all 31 diagrams involving the $b \bar{b} W^+ W^-$
intermediate state are included. Subsequently, two sampling channels were
constructed to describe the topologies in the first and 13 sampling channels in
the second process of the two described above. The results show that the
sampling procedure scales quite efficiently with the increase in the number of
Feynman diagrams and topologies. Detailed studies show that with inclusion of
the new propagator modeling described in this paper the sampling of angular
distributions becomes a new restraining factor in achieving optimal weight
variances and unweighing efficiencies which might therefore be worth
investigating further.

\section{Conclusion}

In this paper the revised Kajantie-Byckling approach and some improved phase
space sampling techniques for the massive multi-particle final states were
presented. In order to adapt the procedure to the LHC environment the
modifications necessary for reversing the sampling order were introduced and
new invariant mass sampling methods, which attempt to describe the propagator
dependence of the probability density together with the phase space suppression
due to the presence of massive particles, were developed.

The developed procedures have been implemented in the {\tt AcerMC 2.0}
Monte--Carlo generator\cite{acermc2}. Basing on the encouraging evidence provided
by the {\tt AcerMC 2.0} implementation of the approach it seems reasonable to
argue that the methods presented in this paper should substantially simplify and
automatise the phase space integration (sampling) techniques while retaining a
respectable weight variance reduction and unweighing efficiencies provided by
the most advanced phase space sampling techniques developed so far
\cite{fermisv,excal,nexcal,bhlumi,acermc}. Furthermore, the authors believe that
these techniques could easily be combined with algorithms of the type {\tt
Sarge} \cite{sarge} or {\tt HAAG} \cite{haag} to provide successful sampling of
the ($\sim$massless) final state particles as e.g. final state gluon radiation.

\section*{Acknowledgments}

This work was inspired by the framework of the Physics Working Groups of the ATLAS
Collaboration. B.P.K. and E.R-W. are grateful to all colleagues for the very
creative atmosphere and several valuable discussions. B.P.K. would like to thank Svjetlana
Fajfer for pointing out several issues of significant relevance for this paper.  



\begin{thebibliography}{99}

\bibitem{jadach}
S. Jadach, {\tt e-print:}{\bf physics/9906056}, (1999). 

\bibitem{was}
 M. Skrzypek, Z. Was,  Comput. Phys. Commun. {\bf 125} (2000) 8.

\bibitem{rambo}
W.J. Stirling, R. Kleiss, and S.D. Ellis, Comput. Phys. Commun. {\bf 40} (1986) 359.

\bibitem{sarge}
 P.D. Draggiotis, A. van Hameren and R. Kleiss,  Phys.Lett. {\bf B483} (2000) 124.

\bibitem{haag}
A. van Hameren and C. G. Papadopoulos, Eur.Phys.J.{\bf C25} (2002), 563.

\bibitem{madgraph}
T. Stelzer and W. F. Long, Comput. Phys. Commun. {\bf 81} (1994) 357. 

\bibitem{kleiss}
R. Kleiss and R. Pittau, Comput. Phys. Commun. {\bf 83} (1994) 141.

\bibitem{vegas}
 G.P. Lepage, J. Comput. Phys. {\bf 27} (1978) 192.

\bibitem{foam}
S. Jadach, Comput. Phys. Commun.  {\bf 130} (2000)  244.

\bibitem{acermc}
B. Kersevan and E. Richter-Was, Comp. Phys. Commun. {\bf 149} (2003) 142.\\
The latest version of the manual and code can be found at:\\
{\tt http://cern.ch/Borut.Kersevan/AcerMC.Welcome.html}

\bibitem{acermc2}
B. Kersevan and E. Richter-Was, {\tt preprint:}{\bf TPJU-6/2004,}\\
{\tt e-print:}{\bf hep-ph/0405247}. 
The latest version of the manual and code can be found at:\\
{\tt http://cern.ch/Borut.Kersevan/AcerMC.Welcome.html}

\bibitem{fermisv}
J. Hilgart, R. Kleiss, F. Le Dibider, Comp. Phys. Comm. {\bf 75} (1993) 191.

\bibitem{excal}
 F.A. Berends, R. Kleiss, Comput. Phys. Commun. {\bf 85} (1996) 11.

\bibitem{tauola}
S. Jadach, J. H. Kuhn, Z. Was, Comput. Phys. Commun. {\bf 64} (1990) 275;
M. Jezabek, Z. Was, S. Jadach, J. H. Kuhn, Comput. Phys. Commun. {\bf 70}
(1992) 69; R. Decker, S. Jadach, J. H. Kuhn, Z. Was, Comput. Phys. Commun. {\bf 76} 
(1993) 361.  

\bibitem{KB}
E. Byckling and K. Kajantie, Nucl. Phys. {\bf B9} (1969) 568.\\
E. Byckling and K. Kajantie, ``{\it Particle Kinematics}'', Wiley \& Sons. , London (1973) 328p. 

\bibitem{Nyborg65a}
P. Nyborg, H.S. Song {\it et al.}, Phys. Rev. {\bf 140}, B914, (1965). 

\bibitem{sampler}
Everett, C J; Cashwell, Edmond D, {\bf LA-9721-MS} (1983) 150p.

\bibitem{lichard}
P. Lichard, Acta Phys.Slov. {\bf 49} (1999) 215.

\bibitem{carlson}
B. C. Carlson, {\bf math.CA/9409227}.

\bibitem{appell}
 P. Appell and J. Kamp\' e de F\' eriet, ``{\it Fonctions hyperg\' eom\' etriques et hypersph\' eriques: polynomes d'Hermite}'', Gauthier-Villars, Paris, 1926.

\bibitem{quad}
Daniel Zwillinger, editor, ``{\it CRC Standard Mathematical Tables and Formulae}'', CRC Press, 30th Edition, (2000).\\
{\it code derived from http://www.csit.fsu.edu/$\sim$burkardt/f\_src/quadrule/quadrule.f90}

\bibitem{numrec}
W.H. Press {\it et al.}, ``{\it Numerical Recipes in Fortran 77}'',  Cambridge Univ. Press, second edition, (1992).

\bibitem{pythia}
T. Sj\" ostrand {\it et al.}, Computer Phys. Commun. {\bf 135} (2001) 238.

\bibitem{nexcal}
F. A. Berends, C. G. Papadopoulos and R. Pittau, Comput.Phys.Commun {\bf 136} (2001) 148.

\bibitem{bhlumi} 
S. Jadach, E. Richter-Was, B.F.L. Ward and Z Was,  Comput. Phys. Commun. {\bf 70} (1992) 305.


\end{thebibliography}
\end{document}